\documentclass[12pt]{article}
\usepackage[numbers]{natbib} 
\usepackage{times}

\newenvironment{sciabstract}{\begin{quote} \bf}
{\end{quote}}

\usepackage[utf8]{inputenc}
\usepackage{soul}
\usepackage{xcolor}
\usepackage[hyphens]{url}
\usepackage[hidelinks]{hyperref}
\usepackage[multiple]{footmisc}
\usepackage{tcolorbox}
\usepackage{subcaption}

\usepackage{soul}
\usepackage{xcolor}
\usepackage{subcaption}
\usepackage{amsmath}
\usepackage{amssymb}
\usepackage{amsthm}

\usepackage{booktabs}
\usepackage{csquotes}
\usepackage{comment}
\usepackage{enumitem}
\usepackage{multirow}
\usepackage{tikz}
\usepackage{array}

\definecolor{L}{HTML}{AB0028}
\definecolor{N}{HTML}{C9BE77}
\definecolor{C}{HTML}{2B3E95}
\definecolor{D}{HTML}{004DFF}
\definecolor{N}{HTML}{C9BE77}
\definecolor{R}{HTML}{FF0000}
\definecolor{LAB1}{HTML}{DF2531}
\definecolor{LAB2}{HTML}{F88082}
\definecolor{RCH}{HTML}{EA77C0}
\definecolor{B60}{HTML}{7030A0}
\definecolor{TVT}{HTML}{FF7D18}
\definecolor{SNP}{HTML}{FFFF00}
\definecolor{ASE}{HTML}{38C177}
\definecolor{SNPO}{HTML}{BEBD3A}
\definecolor{BRX}{HTML}{3AD0DB}
\definecolor{CON}{HTML}{0070C0}

\definecolor{lightyellow}{cmyk}{0,0,0.50,0}
\sethlcolor{lightyellow}

\let\oldFootnote\footnote
\newcommand\nextToken\relax

\renewcommand\footnote[1]{\oldFootnote{#1}\futurelet\nextToken\isFootnote}

\newcommand\isFootnote{\ifx\footnote\nextToken\textsuperscript{,}\fi}

\title{Temporal Dynamics of Coordinated Online Behavior: Stability, Archetypes, and Influence\thanks{\textcolor{red}{Article published in the \textit{Proceedings of the National Academy of Sciences 121(20) -- PNAS}. DOI: \href{http://doi.org/10.1073/pnas.2307038121}{10.1073/pnas.2307038121}. Please, cite the published version.}}}

\author
{Serena Tardelli,$^{1}$  Leonardo Nizzoli,$^{1}$  Maurizio Tesconi,$^{1}$\\ Mauro Conti,$^{2}$  Preslav Nakov,$^{3}$  Giovanni Da San Martino,$^{2}$  Stefano Cresci,$^{{1}\dagger}$\\
\\
\normalsize{$^{1}$IIT-CNR, Italy}; \normalsize{$^{2}$University of Padua, Italy}; \normalsize{$^{3}$MBZUAI, UAE}\\
\\
\normalsize{$^\dagger$To whom correspondence should be addressed; E-mail: stefano.cresci@iit.cnr.it}
}
\date{}
\begin{document} 

\baselineskip18pt
\maketitle 

\begin{sciabstract}
Large-scale online campaigns, malicious or otherwise, require a significant degree of coordination among participants, which sparked interest in the study of \textit{coordinated online behavior}. State-of-the-art methods for detecting coordinated behavior perform \textit{static} analyses, disregarding the temporal dynamics of coordination. Here, we carry out the first \textit{dynamic} analysis of coordinated behavior. To reach our goal we build a multiplex temporal network and we perform dynamic community detection to identify groups of users that exhibited coordinated behaviors in time. We find that: \textit{(i)} coordinated communities feature variable degrees of temporal instability; \textit{(ii)} dynamic analyses are needed to account for such instability, and results of static analyses can be unreliable and scarcely representative of unstable communities; \textit{(iii)} some users exhibit distinct archetypal behaviors that have important practical implications; \textit{(iv)} content and network characteristics contribute to explaining why users leave and join coordinated communities. Our results demonstrate the advantages of dynamic analyses and open up new directions of research on the unfolding of online debates, on the strategies of coordinated communities, and on the patterns of online influence.
\end{sciabstract}

\section*{Introduction}
Online platforms offer unprecedented opportunities to organize and carry out large-scale activities, defying the constraints that limit physical interactions. A wide range of activities benefit from such opportunities, including legitimate information campaigns and political protests \mbox{\cite{gonzalez2011dynamics,steinert2015online,francois2021measuring}}, as well as potentially nefarious ones such as disinformation campaigns and targeted harassment~\cite{lazer2018science,starbird2019disinformation,bovet2019influence,mariconti2019you,dipietro2021new}. A common characteristic of large-scale online campaigns is the significant degree of \textit{coordination} among the  users involved, which is needed to spread content widely and to let the campaigns obtain significant outreach, ultimately ensuring their success~\cite{nizzoli2021coordinated}.

Due to the pervasiveness of coordinated online behavior and its relevance for the effectiveness of both benign and malicious campaigns, the topic gained much scholarly attention. For example, a few methods were recently proposed for detecting coordinated communities (CCs) and for measuring the extent of coordination between users~\cite{pacheco2020uncovering,nizzoli2021coordinated,weber2021amplifying}. These studies define coordination as an unexpected or exceptional similarity between the actions of two or more users. The analysis is carried out by building a user similarity network based on common user activities (e.g., co-retweets) and by studying it with network science techniques. A limitation of the existing methods is that they are based on \textit{static} analyses. For example,~\cite{nizzoli2021coordinated,pacheco2020uncovering,weber2021amplifying} each build a single aggregated network that encodes user behaviors occurred throughout many weeks. However, online behaviors are \textit{dynamic} (i.e., time-varying)~\cite{sekara2016fundamental,weber2021temporal}. As such, aggregating behaviors over many weeks (or months) of time represents an oversimplification that risks overshadowing important temporal dynamics. On the contrary, a minority of detection methods solely model time~\cite{sharma2020identifying,zhang2021vigdet}, disregarding other important facets of online coordination such as the interactions between users.  Due to the above methodological limitations, the temporal dynamics of coordinated behavior are, so far, essentially unexplored.
In addition, the majority of the existing literature on coordinated behavior, especially from the area of computer science, focus on \textit{inorganic} and \textit{malicious} coordination. This is because of the nefarious consequences that such forms of coordination have on the online environments and even on the society at large, which make inorganic and malicious coordination worthy of particular attention. However, both foundational works on \textit{offline} coordination~\mbox{\cite{malone1994interdisciplinary}} as well as more recent works on online coordination~\mbox{\cite{giglietto2020takes,magelinski2022synchronized,nizzoli2021coordinated}} remark that coordination can also be \textit{implicit} rather than explicit, and \textit{spontaneous} and \textit{emergent} rather than intentional, inorganic, and organized. In light of this literature, here we refer to coordinated behavior in an intentionally broad and unbiased way. Our choice to encompass all instances of online coordination -- malicious and inorganic, or otherwise -- allows us to investigate both harmful coordinated groups and neutral ones. Therefore, in addition to providing contributions to the study of online information manipulation, our work also contributes to the understanding of human dynamics and online interactions, providing interesting findings from both computational and social standpoints.

\subsection*{Contributions}
We carry out the first dynamic analysis of coordinated behavior. Instead of working with static networks, we build and analyze a \textit{multiplex temporal network}~\mbox{\cite{mucha2010community}} and we leverage state-of-the-art dynamic community detection algorithms~\mbox{\cite{rossetti2018community}} to find groups of users that exhibited coordinated behaviors in time. We apply our method to two Twitter datasets of politically-polarized discussions covering the run up to the 2019 UK general elections~\mbox{\cite{nizzoli2021coordinated,hristakieva2022spread}} and the 2020 USA presidential elections. In addition, we validate our method on a third Twitter dataset featuring known instances of malicious accounts involved in a large information operation (See \textit{Materials and Methods} for details on the datasets and our methodology). We compare the results of our novel dynamic analysis to the static ones, demonstrating the advantages of the former. Most importantly however, our innovative approach opens up new directions of research and allows answering to the following research questions:
\subsubsection*{RQ1}
\textit{To what extent are coordinated communities stable over time?} So far, coordinated behavior has been studied with static methods. However, some CCs might exhibit temporal instability, meaning that, through time, a significant portion of members leave the community while others join it. In this case, time-aggregated static analyses would yield unreliable results by overshadowing the temporal variations. Here, we compare static and dynamic results, and we shed light on the stability of CCs. \subsubsection*{RQ2}
\textit{What are the temporal dynamics of user behavior with respect to the evolving coordinated communities?} In case a certain degree of instability existed, some users would belong to different CCs at different points in time. Here, we investigate the dynamics with which user membership to CCs changes through time and their implications on the effectiveness of online campaigns.
\subsubsection*{RQ3}
\textit{Why do users belong to different coordinated communities through time?} Different patterns of user membership to CCs are indicative of markedly different situations. For instance, users who remain in the same community for a long time might be strong supporters of that community. Conversely, users who abandon a community in favor of another might have been disappointed by the former or persuaded by the latter. In any case, investigating the possible reasons for user shifts between CCs (or lack thereof) is a novel direction of research with important practical implications (e.g., the study of online influence). Here, we analyze users who exhibit different behaviors and we compare them to their respective CCs, gaining insights into why users change community.

Based on the results to the aforementioned RQs, our main contributions are summarized as follows:
\begin{itemize}
    \item We carry out the first dynamic analysis of coordinated online behavior.
    \item We show that the communities involved in both the UK 2019 and the USA 2020 electoral debates featured variable degrees of \textit{instability}, which motivates dynamic analyses.
    \item We find that the majority of user shifts from a community to another occurred between \textit{similar} (like-minded) communities, while only a minority involved very different communities.
    \item We define and characterize three archetypes of users with markedly different behaviors, namely: \textit{(i)} \textit{stationary}, \textit{(ii)} \textit{influenced}, and \textit{(iii)} \textit{volatile} users.
    \item We find that \textit{content} and \textit{network} characteristics are useful for understanding why users move between communities.
\end{itemize}

\subsubsection*{Significance}
Social media campaigns shape public opinion in various domains of society. Coordinated behavior is at the heart of successful campaigns, enabling outreach through effective content spreading. This study focuses on the temporal dynamics of coordinated campaigns and characterizes their influence on Twitter during two recent major elections. Analyzing the temporal changes of online behavior uncovers complex dynamic patterns that evolve over time. We reveal different user behaviors and archetypes, measuring how they affect and influence the online environment. Investigating the temporal dimension of online coordination reveals the dynamics of online debates, the strategies of coordinated communities, and the patterns of online influence, with major practical implications for research and policy of online platforms and for the society at large.
 \section*{Related Work}
\label{sec:relatedwork}
The majority of existing approaches for detecting coordinated behavior are based on network science. These works model common activities between users (e.g., co-retweets, temporal and linguistic similarities, etc.) to build user similarity networks and to subsequently analyze them, for example by means of community detection algorithms~\cite{weber2021amplifying,pacheco2020uncovering,magelinski2022synchronized}. The typical output of these methods is a network where the CCs are identified. Some network-based methods do not only detect coordination as in a binary classification task, but also quantify the extent of coordination between users and communities, thus providing more nuanced results~\cite{nizzoli2021coordinated}. Notably, all aforementioned methods build static networks and employ static community detection algorithms. 
A few works focused on describing and even predicting temporal changes in online community structure. However, these leveraged explicit communities, such as the scholars belonging to a specific scientific community~\mbox{\cite{backstrom2006group}}, or the relationship between gamers and teams in certain team-based online games~\mbox{\cite{patil2013predicting}}. In our work instead, the online communities are not known in advance. Other significant differences with respect to the existing works are related to the temporal granularity of the analysis. For example, some studies measured user migrations between platforms after major events, such as the bans of toxic communities from a certain platform~\mbox{\cite{newell2016user,horta2021platform}}. However, such analyses only considered before/after scenarios, without providing a fine-grained temporal modeling of the user migrations nor a nuanced network representation of the communities involved in the event.
So far, the dynamic analysis of coordinated behavior is essentially unexplored. To this end however, several advances were proposed to model time-varying behaviors with multiplex temporal networks~\cite{mucha2010community} and to employ dynamic community detection algorithms~\cite{rossetti2018community}. Our work applies these techniques -- for the first time -- to the study of coordinated behavior, thus moving beyond the current state-of-the-art. In addition to network-based approaches, others proposed to detect coordination with temporal point processes where user activities are modeled as the realization of a stochastic process~\cite{sharma2020identifying,zhang2021vigdet}. These methods are capable of modeling the latent influence between the coordinated accounts, their strongly organized nature, and possible prior available knowledge. Finally, others adopted traditional feature engineering approaches to find similarities between users~\cite{francois2021measuring}, or focused on specific user behaviors such as URL sharing~\cite{giglietto2020takes}.

Once coordinated behaviors are detected, subsequent efforts are devoted to characterizing CCs. Characterization tasks are typically aimed at distinguishing between malicious (e.g., disinformation networks) and genuine (e.g., fandoms, activists) forms of coordination~\cite{vargas2020detection,mendoza2020bots,hristakieva2022spread,simchon2022troll}. This can be achieved by analyzing the content shared by the CCs, as done by~\cite{hristakieva2022spread} that estimated the amount of shared propaganda. Others analyzed the structural properties of the coordination networks, finding differences between malicious and benign CCs~\cite{vargas2020detection,nizzoli2021coordinated}. Finally, another interesting direction of research revolves around estimating the influence that CCs have on other users. However, existing results in this area are still scant and contradictory. For example,~\cite{cinelli2022coordinated} studied network properties of information cascades on Twitter, finding that coordinated users have a strong influence on the non-coordinated ones that participate in the same cascade. Contrarily,~\cite{sharma2020identifying} found that coordinated users have a strong influence on other coordinated users and only a small influence on non-coordinated ones.
Here, we show that temporal coordination networks are a valuable tool towards assessing the influence that CCs exert on the users in a network.
 \section*{Results}
\label{sec:results}
The application of our method (See \textit{Materials and Methods} for a detailed discussion of our methodological approach) to the UK 2019 electoral debate produced a multiplex temporal network $G$ comprising 277K nodes and 4.1M edges. On average, a layer $G_i \in G$ contains 11K nodes ($\sigma = 169$) and 164K edges ($\sigma = 19$K). In total, $\simeq 600$ different CCs were part of $G$ for at least one time window. The analysis of the USA 2020 debate resulted in a multiplex temporal network $G$ comprising 526K nodes and 7.5M edges. On average, a layer $G_i \in G$ contains 21K nodes ($\sigma = 2$K) and 303K edges ($\sigma = 79$K). In total, $\simeq 2$K different CCs were part of $G$ for at least one time window. Out of all the CCs identified in both the UK 2019 and the USA 2020 datasets, only a limited number persisted for all of the time. These also include the vast majority of all users in the networks. For this reason, in the following we present detailed results only for the top-10 largest CCs from both datasets (See \textit{Supporting Table}~S3).

\subsection*{Dynamic \textit{vs.} static communities}
We first present the results of our dynamic analyses and we subsequently compare them with the static ones.

\subsubsection*{Dynamic UK 2019 communities}
The dynamic communities that dominated the 2019 UK electoral debate are as follows:
\begin{itemize}
    \item \texttt{LAB1}: Labourists that supported the party and its leader Jeremy Corbyn, in addition to traditional Labour themes such as healthcare and climate change.
    \item \texttt{LAB2}: Another community of labourists. Their themes were similar to those of \texttt{LAB1}, but their members exhibited different temporal behaviors (See \textit{Supporting Figure}~S1).
    \item \texttt{RCH}: Users who spread the labourist manifesto and urged others to register for the vote. \texttt{RCH} is mostly similar to \texttt{LAB2} but, again, featured different temporal behaviors (See \textit{Supporting Figure}~S1).
    \item \texttt{B60}: Activists that used the electoral debate to protest against a pension age equalisation law to the detriment of women born in the 1950s~\cite{backto60}.\item \texttt{TVT}: A community comprising multiple political parties that pushed for a tactical vote in favor of the labourists, to oppose the conservative party and to stop the Brexit.
    \item \texttt{SNP}: Supporters of the Scottish National Party (SNP), in favor of the Scottish independence from the UK.
    \item \texttt{ASE}: Conservative supporters that were mainly engaged in attacking the labour party and Jeremy Corbyn for antisemitism~\cite{Revealed}.\item \texttt{SNPO}: Users that debated and opposed Scotland’s intention of pushing for a second independence referendum.
    \item \texttt{BRX}: A community in support of the former Brexit Party.
    \item \texttt{CON}: Conservatives that supported the party and its leader Boris Johnson, as well as Brexit.
\end{itemize}

\begin{figure}[!t]
\centering
    \begin{minipage}{.6\columnwidth}\includegraphics[width=\textwidth]{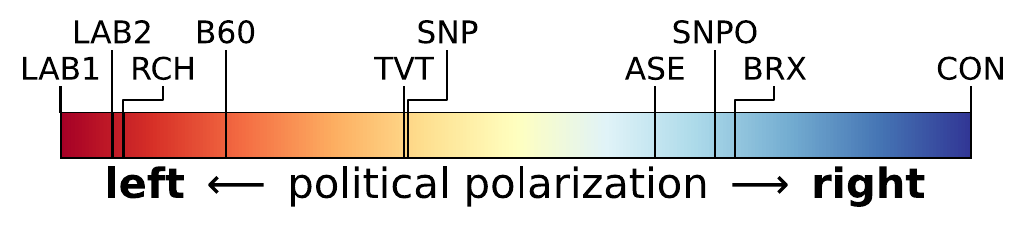}
        \subcaption{UK 2019.}
        \label{fig:polarity-spectrum-uk}
    \end{minipage}\hspace{.05\textwidth}\begin{minipage}{.6\columnwidth}\includegraphics[width=\textwidth]{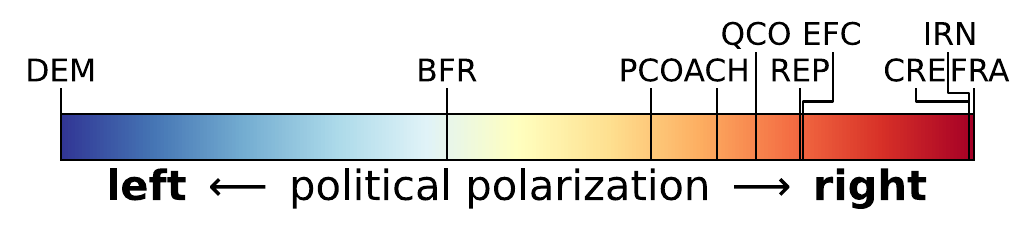}
        \subcaption{USA 2020.}
        \label{fig:polarity-spectrum-us}
    \end{minipage}\caption{Position of the CCs in the political spectrum. The color scheme mirrors the political or ideological colors of the country.}
    \label{fig:polarity-spectrum}
\end{figure}

Adding to the content analysis of the themes discussed by each community, we also characterize CCs by their political orientation (See \textit{Materials and Methods}). Figure~\ref{fig:polarity-spectrum-uk} shows the position of each CC within the continuous political spectrum. Overall, the dynamic CCs identified with our method are in line with the 2019 UK political landscape and with the development of the electoral debate~\cite{jackson2019uk}. CCs represent both large and strongly polarized parties (e.g., labourists and conservatives), as well as smaller ones that teamed up against conservatives with tactical voting. As shown in Figure~\ref{fig:polarity-spectrum-uk}, each CC also appears to be correctly positioned within the 2019 UK political spectrum, with \texttt{LAB1}, \texttt{LAB2}, and \texttt{CON} holding the extremes of the spectrum, and parties such as liberal democrats and Scottish nationalists (\texttt{TVT} and \texttt{SNP}, respectively) laying towards the middle.

\subsubsection*{Dynamic USA 2020 communities}
We again characterize the top dynamic communities by analyzing their use of hashtags through time (See \textit{Supporting Figure}~S1). The dynamic communities that dominated the 2020 USA electoral debate are as follows: \begin{itemize}
    \item \texttt{DEM}: Democrats that supported the party and its leader Joe Biden.
    \item \texttt{BFR}: A community engaged in promoting Biafra's independence from Nigeria while also supporting Trump\mbox{~\cite{Niger}}.\item \texttt{PCO}: Users engaged in Pandemic conspiracy theories and other controversial narratives about the China government responsibility in the COVID-19 pandemic.
    \item \texttt{QCO}: Users engaged in the QAnon conspiracy theory\mbox{~\cite{qanon}}.\item \texttt{ACH}: Hong Kong protesters against the China's regime and in support of Trump\mbox{~\cite{hongkong}}.\item \texttt{EFC}: Users discussing the allegation of election fraud due to postal ballots.
    \item \texttt{REP}: Republicans that supported the party and its leader Donald Trump.
    \item \texttt{FRA}: Far-right French users in support of Trump and the voting fraud accusations.
    \item \texttt{CRE}: A community in support of the Republican party, engaging in conspiracy theories.
    \item \texttt{IRN}: Users in support of the \textit{Restart} movement, an Iranian political opposition group, which supports Trump for his policy against the Iranian regime\mbox{~\cite{iran}}.\end{itemize}

We compute the political polarization of each CC with the same technique adopted for the UK 2019 CCs (See \textit{Materials and Methods}). Figure~{\ref{fig:polarity-spectrum-us}} shows the position of each CC within the continuous political spectrum. The dynamic CCs identified with our method are consistent with the political environment in the run up to the 2020 USA presidential elections~\mbox{\cite{ferrara2020characterizing}}. The main differences between the results obtained for the USA 2020 versus the UK 2019 dataset are due to the inherently different online political environment at the time of the two elections. Specifically, while the online debate about the UK 2019 political election involved communities scattered throughout the whole political spectrum, as shown in Figure~{\ref{fig:polarity-spectrum-uk}}, the one about the USA 2020 election involved an overwhelming majority of Republican-aligned communities and only an isolated community of Democrat users, as shown in Figure~{\ref{fig:polarity-spectrum-us}}.
This result is also confirmed when analysing the topic-based pairwise similarities between CCs (See \textit{Supporting Figure}~S2).

\begin{figure}[!t]
\centering
    \begin{minipage}{.47\columnwidth}\includegraphics[width=\textwidth]{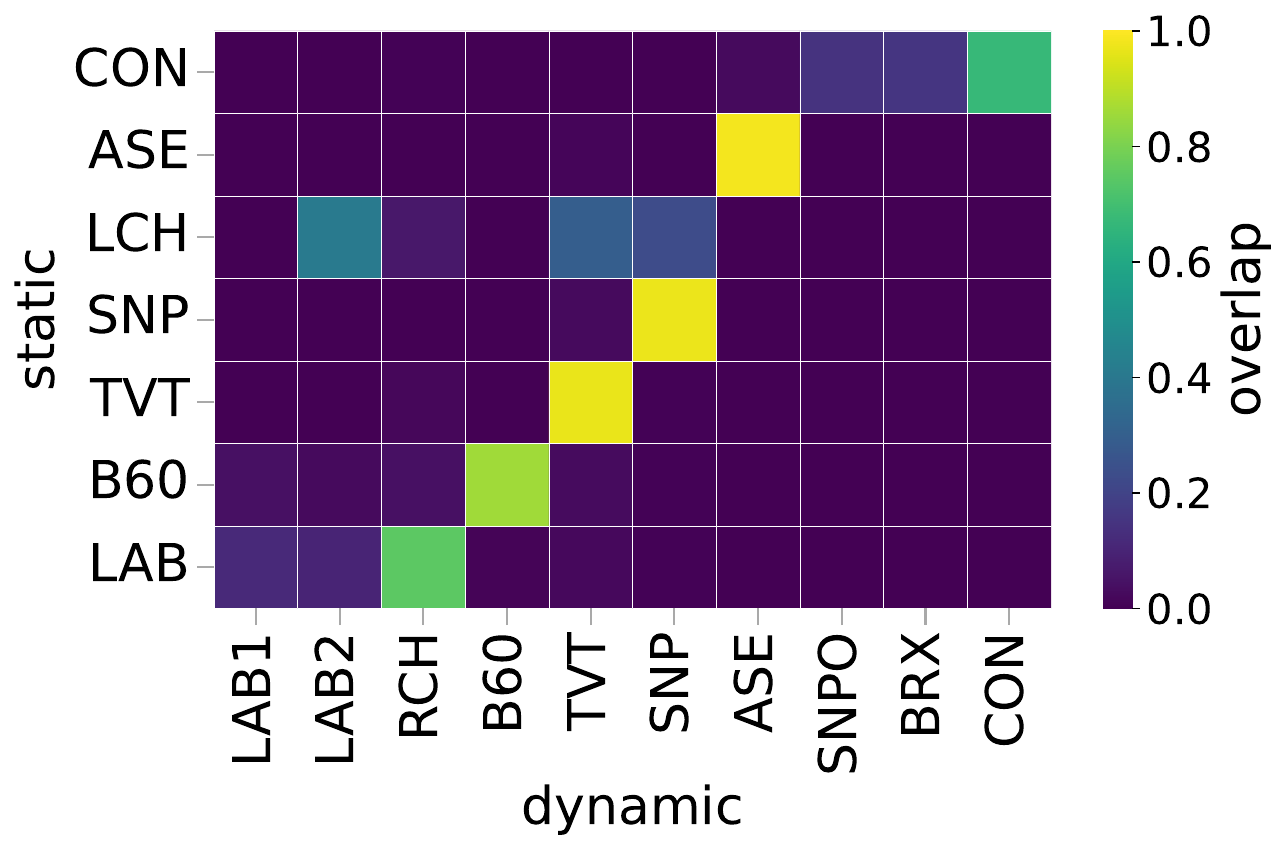}
        \subcaption{UK 2019.}
        \label{fig:static-vs-dynamic-uk}
    \end{minipage}\hspace{.05\columnwidth}\begin{minipage}{.47\columnwidth}\includegraphics[width=\textwidth]{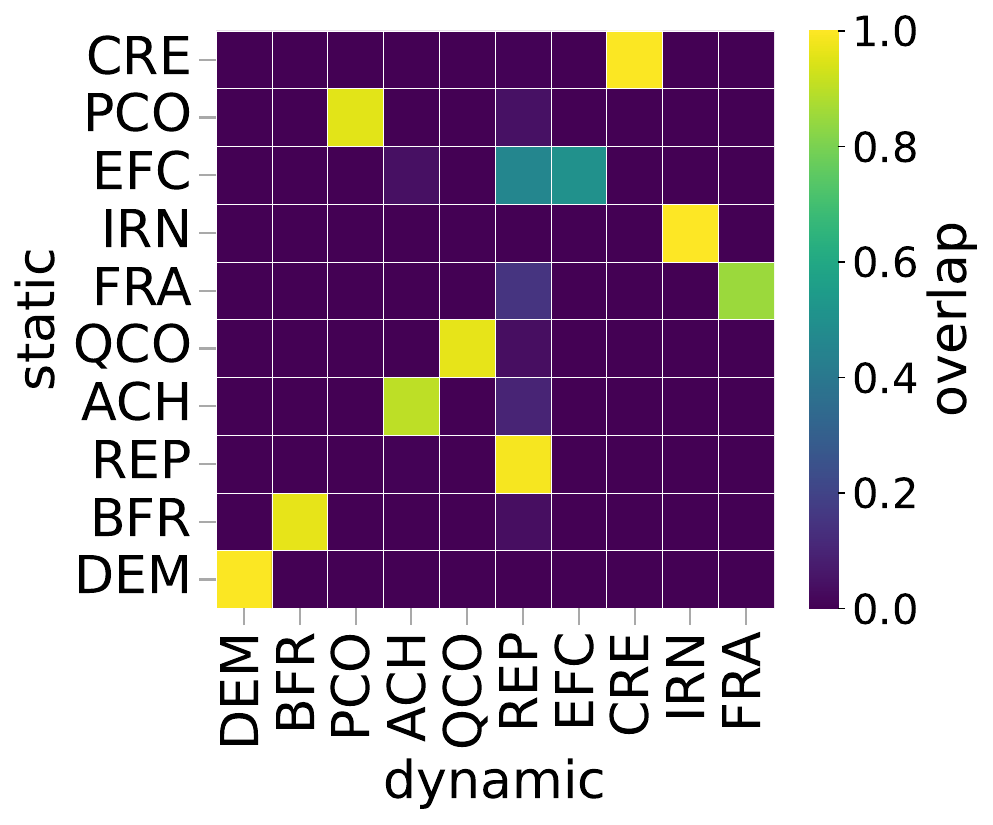}
        \subcaption{USA 2020.}
        \label{fig:static-vs-dynamic-us}
    \end{minipage}\caption{Mapping between the static CCs found in previous studies and the dynamic CCs found with our method. The mapping is based on the overlap between community members. Overall there is a good match between static and dynamic communities.} \label{fig:static-vs-dynamic}
\end{figure}

\subsubsection*{Comparison with static CCs}
Here we compare the results of our dynamic analyses with those obtained with static analyses on each dataset. Previous studies~\mbox{\cite{nizzoli2021coordinated}} found 7 static CCs in the UK 2019 dataset, while we found 10 static CCs in the USA 2020 dataset. Figure~\mbox{\ref{fig:static-vs-dynamic}} presents a heatmap of the overlap between the static CCs (\textit{y} axis) and the dynamic ones (\textit{x} axis) found with our method for each dataset. In figure the overlap expresses the fraction of users from a static CC that are members of a dynamic one. As shown, there is overall a very good match between static and dynamic CCs. Specifically, 5 out of 7 and 9 out of 10 static CCs have overlap $> 60\%$ with a dynamic CC, for UK 2019 and USA 2020, respectively. This result supports the consistency between static and dynamic analyses. Nonetheless, some static communities from each dataset were split into multiple dynamic communities. For UK 2019, the \texttt{CON}, \texttt{LCH} (a small community of activists protesting against an unfair taxation), and \texttt{LAB} static CCs were split, whereas for USA 2020 this occurred to \texttt{EFC}. The results of the dynamic analyses are confirmed by the analysis of the temporal behaviors of these CCs (See \textit{Supporting Figure}~S1). As an example, users of the \texttt{LAB} static CC belong to the \texttt{LAB1}, \texttt{LAB2}, and \texttt{RCH} dynamic CCs, which all exhibit overall similar but temporally different behaviors. To this regard, the larger number of CCs found with the dynamic analysis are due to the better modeling of the temporal dimension, which allows identifying behavioral differences that were not observable with static analyses.

\subsection*{RQ1: Temporal stability of coordinated communities}
So far we highlighted that our dynamic analyses provided overall similar -- yet more nuanced and fine-grained -- results with respect to previous static analyses. Now we leverage our innovative dynamic model of coordinated behavior (See \textit{Materials and Methods}) to evaluate the temporal stability of the CCs involved in the online electoral debates, by investigating the extent to which they underwent changes through time. This analysis is relevant for multiple reasons: \textit{(i)} it sheds light, for the first time, on how CCs evolve in time, and \textit{(ii)} it allows assessing how representative are static analyses of coordinated behavior, with respect to the temporal variations of the CCs.

\begin{figure}[!t]
    \centering
    \begin{minipage}{.3\columnwidth}\includegraphics[width=\textwidth]{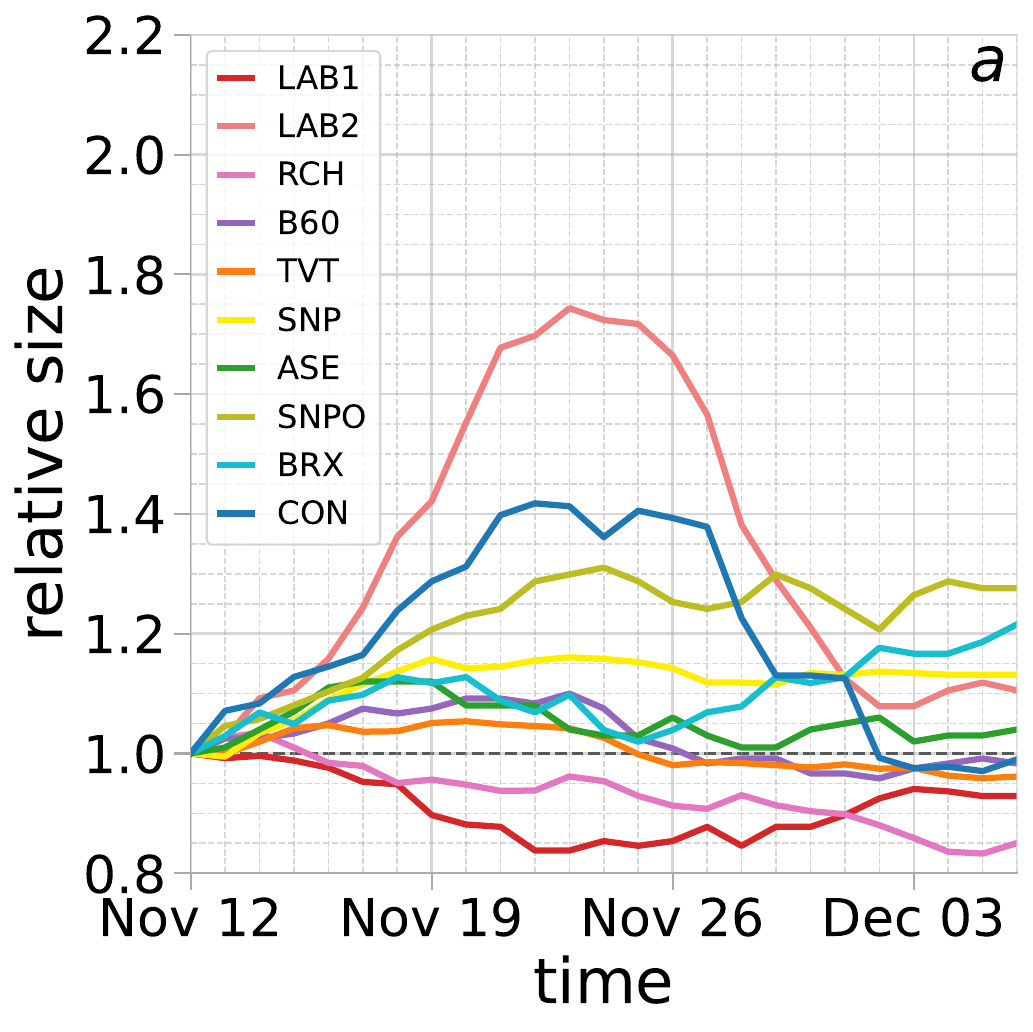}
    \end{minipage}\hspace{.05\columnwidth}\begin{minipage}{.3\columnwidth}\includegraphics[width=\textwidth]{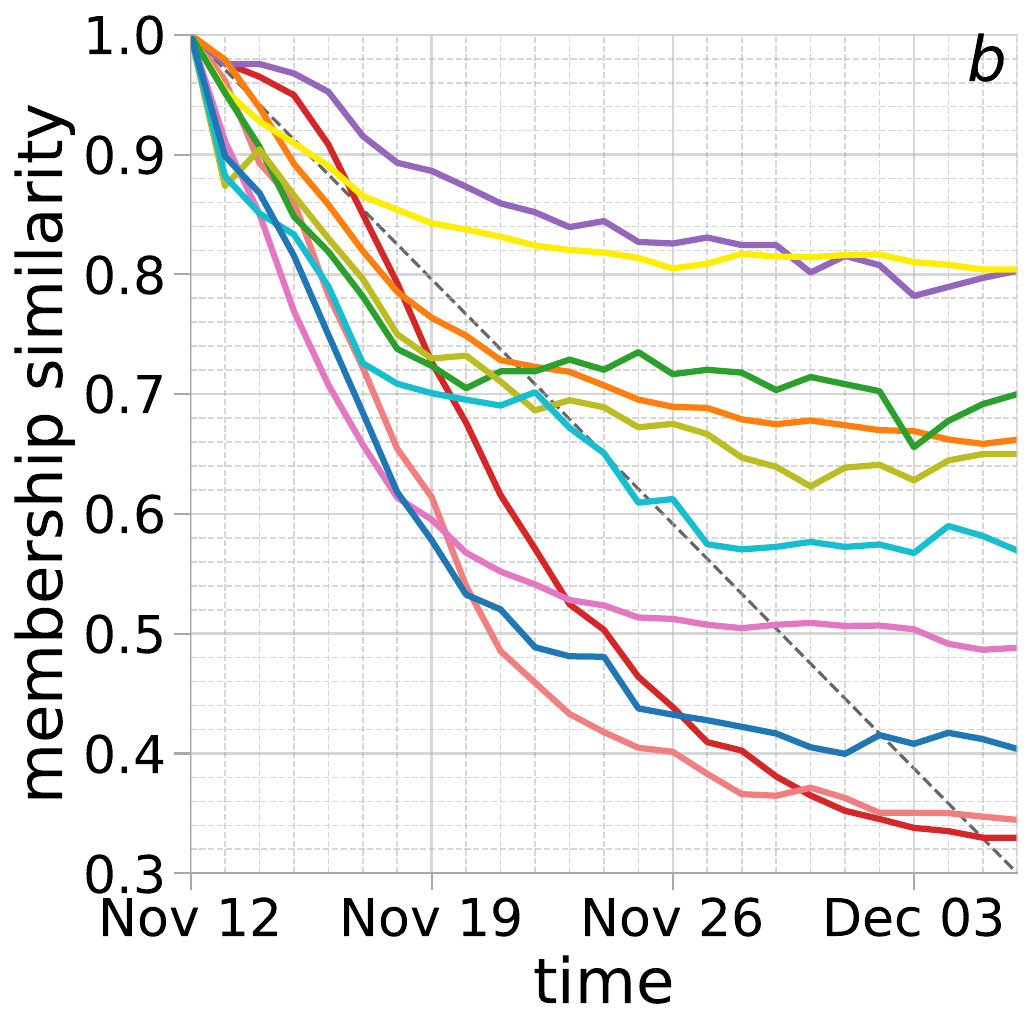}
    \end{minipage}\\
    \begin{minipage}{.3\columnwidth}\includegraphics[width=\textwidth]{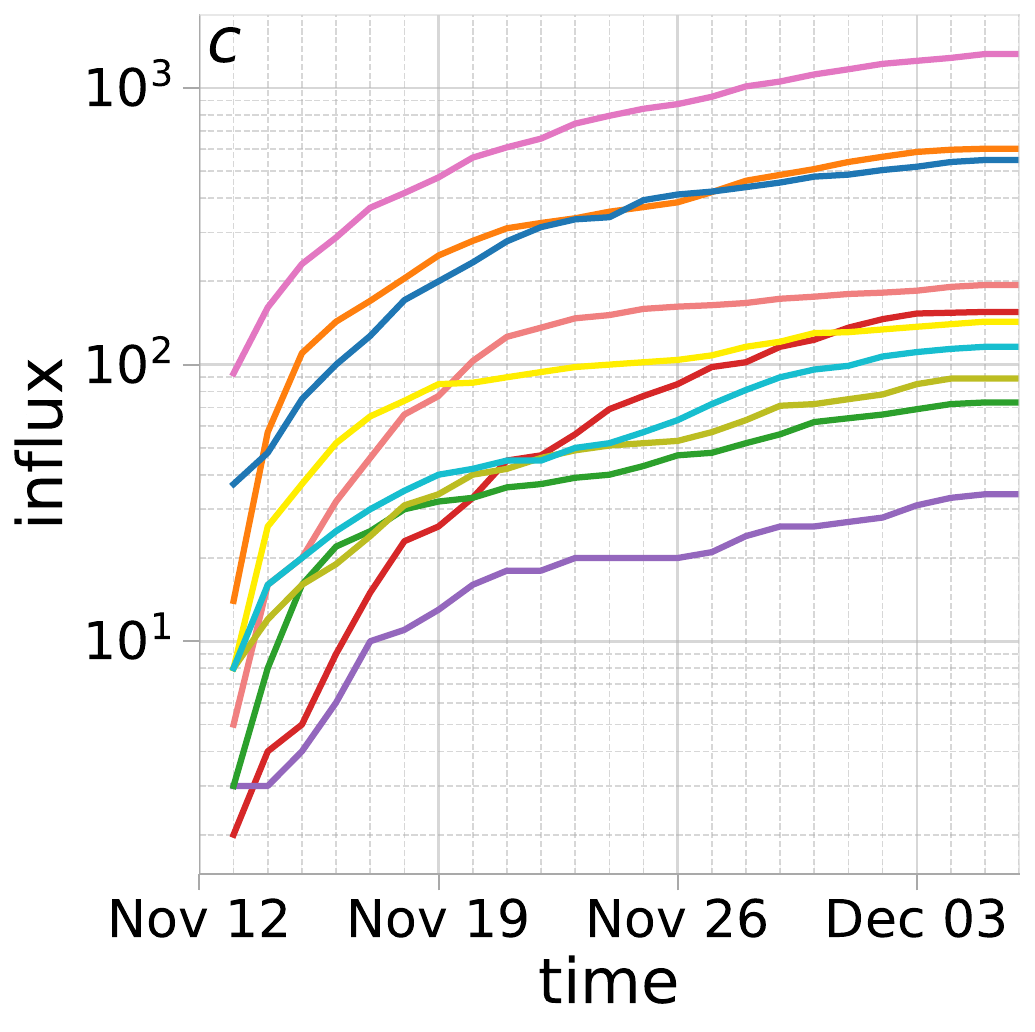}
    \end{minipage}\hspace{.05\columnwidth}\begin{minipage}{.3\columnwidth}\includegraphics[width=\textwidth]{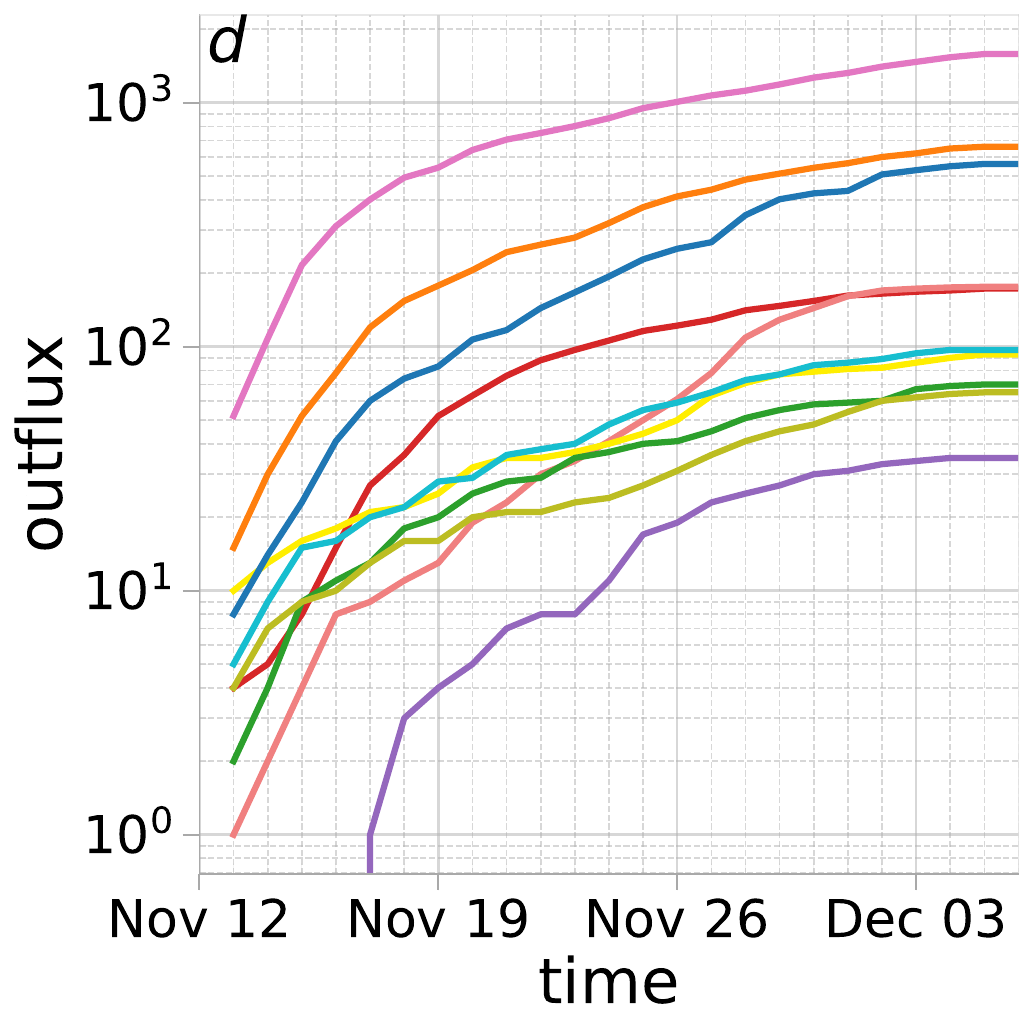}
    \end{minipage}\caption{UK 2019: Temporal stability of the CCs measured in terms of their evolving size (a), membership (b), and influx (c) and outflux (d) of users to/from the community. Each tick on the \textit{x} axis corresponds to a one week-long time window. Time windows are offset by one day. Dates on the \textit{x} axis represent the start date of the corresponding time window.}
    \label{fig:temporal-stability}
\end{figure}

\subsubsection*{Temporal evolution of CCs}
\begin{figure}[!t]
    \centering
    \begin{minipage}{.3\columnwidth}\includegraphics[width=\textwidth]{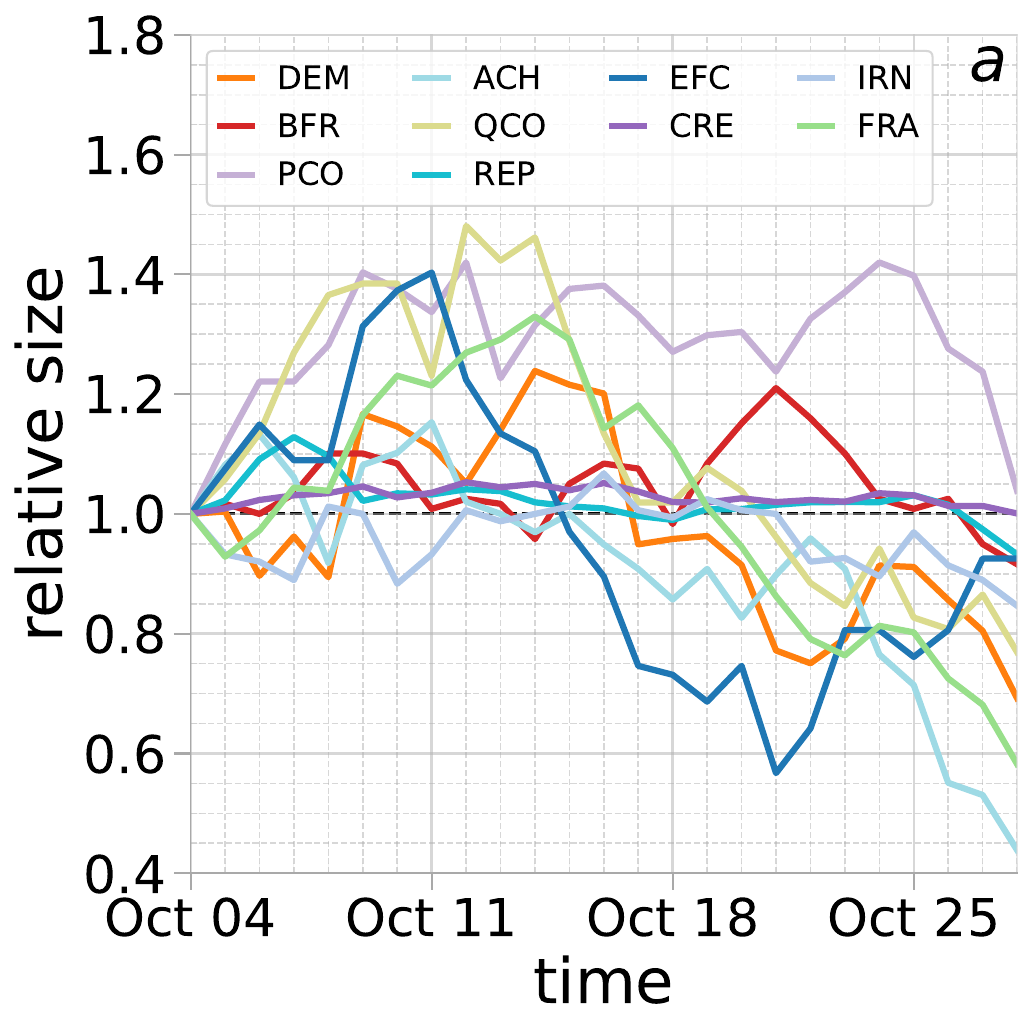}
    \end{minipage}\hspace{.05\columnwidth}\begin{minipage}{.3\columnwidth}\includegraphics[width=\textwidth]{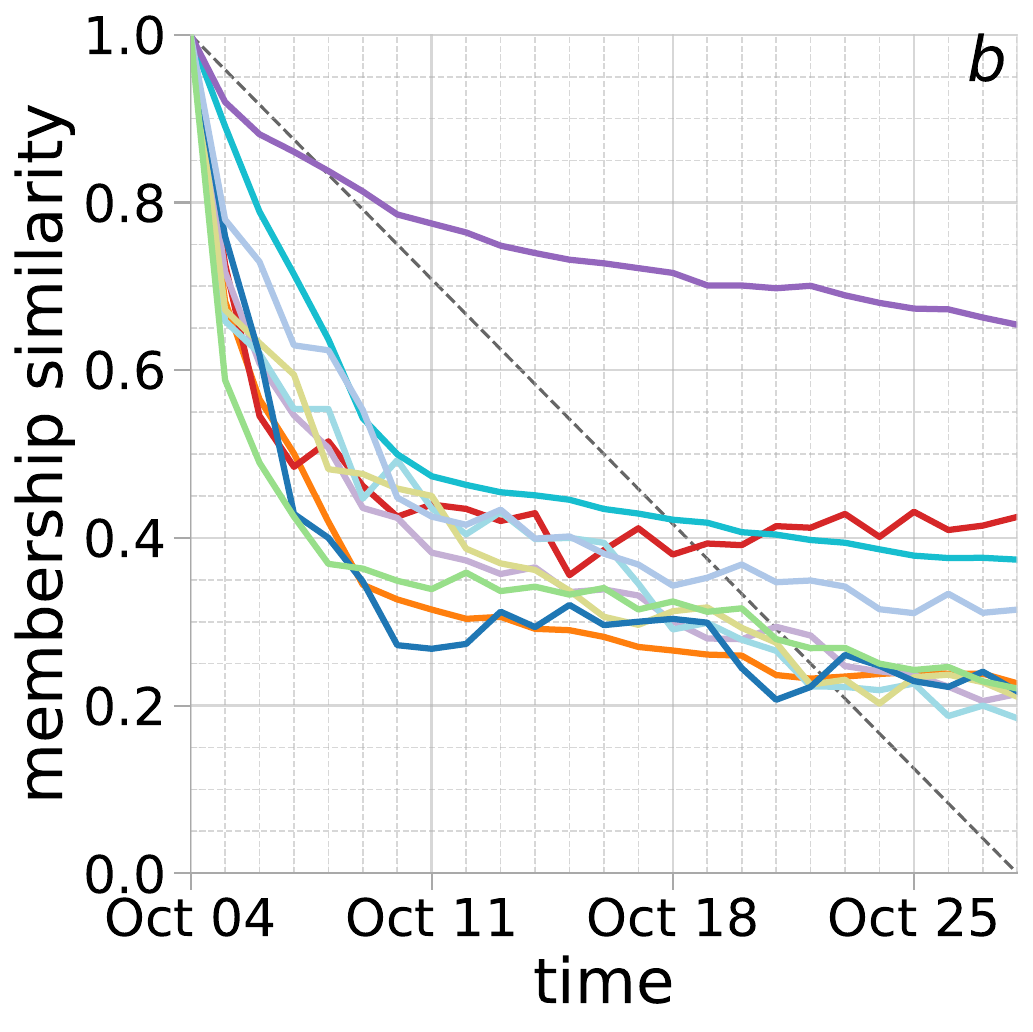}
    \end{minipage}\\
    \begin{minipage}{.3\columnwidth}\includegraphics[width=\textwidth]{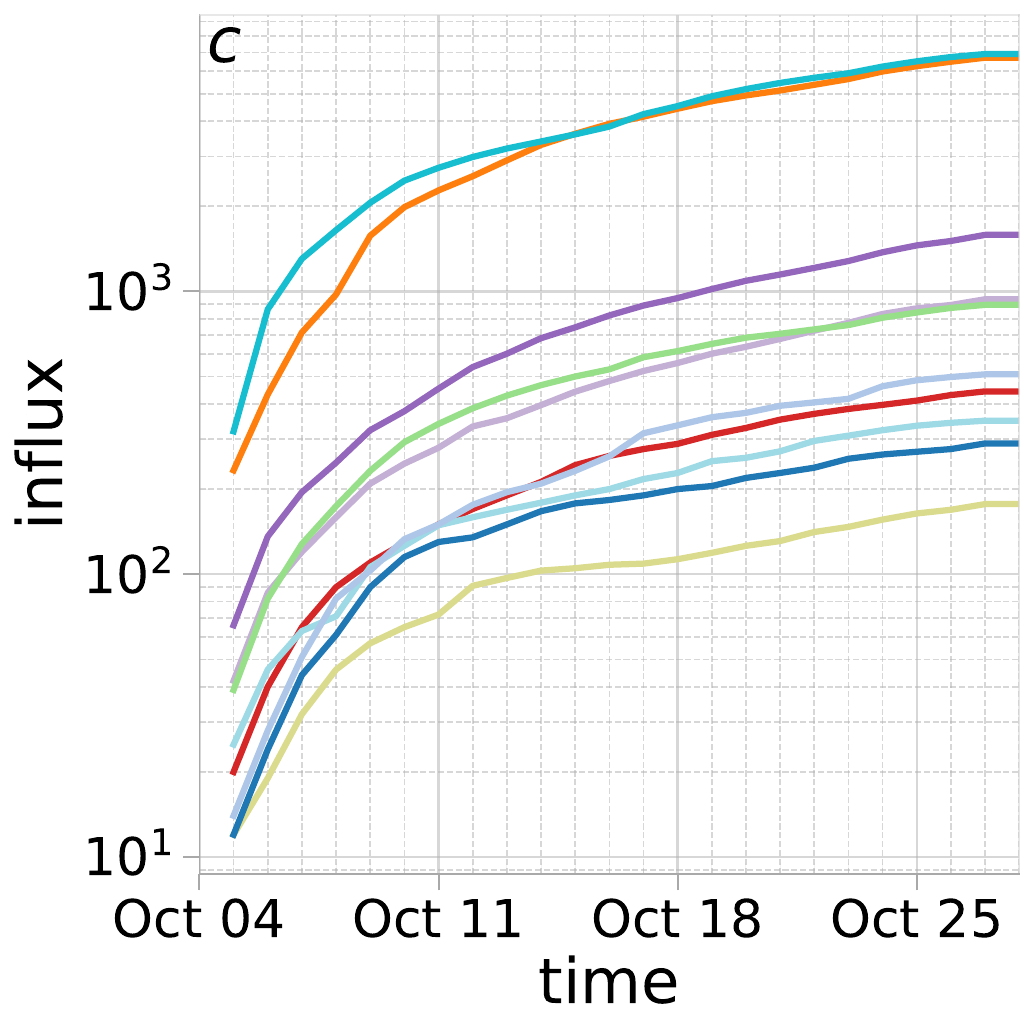}
    \end{minipage}\hspace{.05\columnwidth}\begin{minipage}{.3\columnwidth}\includegraphics[width=\textwidth]{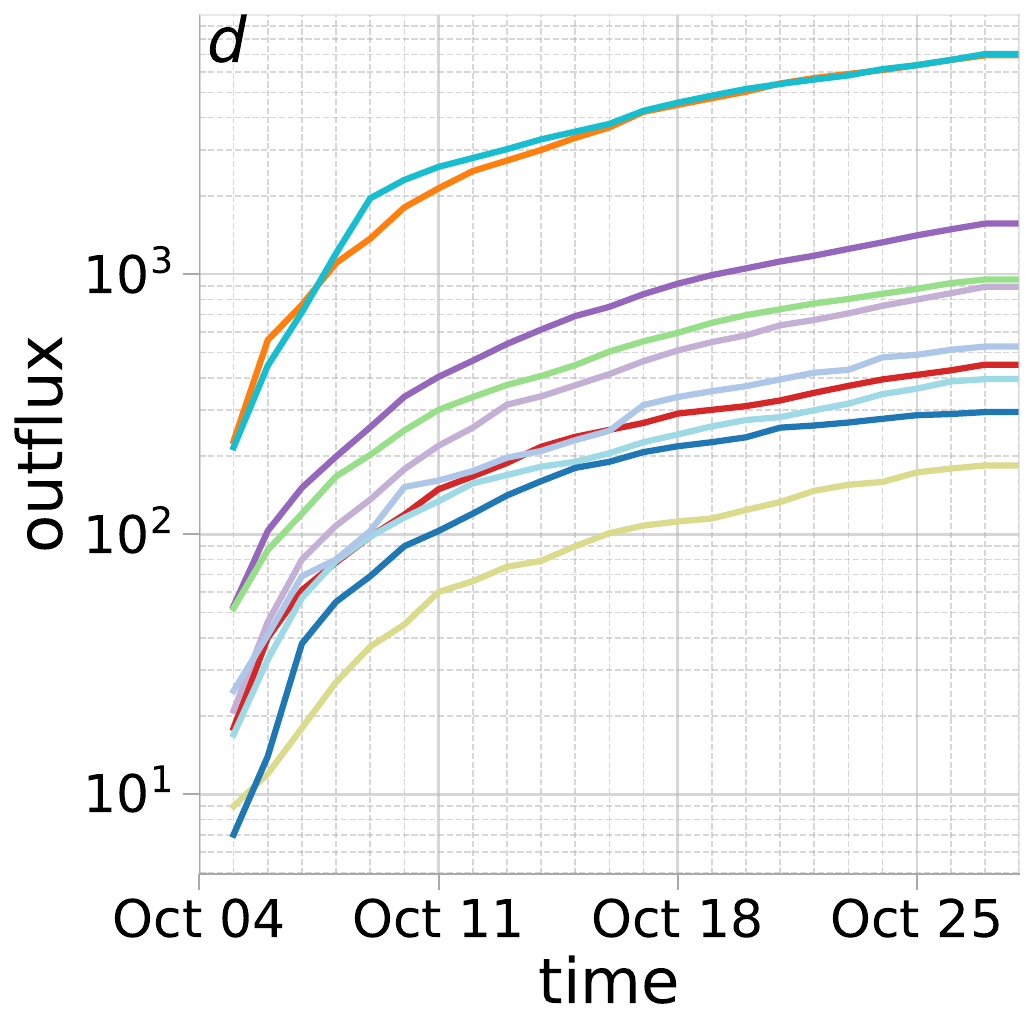}
    \end{minipage}\caption{USA 2020: Temporal stability of the CCs measured in terms of their evolving size (a), membership (b), and influx (c) and outflux (d) of users to/from the community. Each tick on the \textit{x} axis corresponds to a one week-long time window. Time windows are offset by one day. Dates on the \textit{x} axis represent the start date of the corresponding time window.}
    \label{fig:temporal-stability-us}
\end{figure}

To assess how CCs changed through time we measured temporal fluctuations in their size, membership, as well as their cumulative influx and outflux of users (See \textit{Materials and Methods}). Figures~\mbox{\ref{fig:temporal-stability}} and~\mbox{\ref{fig:temporal-stability-us}} present the results of these analyses. Regarding community size, Figures~\mbox{\ref{fig:temporal-stability}}a and and~\mbox{\ref{fig:temporal-stability-us}}a highlight major differences in the temporal evolution of the CCs. Some communities such as \texttt{SNP}, \texttt{TVT}, \texttt{ASE}, \texttt{IRN}, and \texttt{CRE} appear as relatively stable in time, with only limited size fluctuations represented by mostly flat lines. Contrarily, other communities such as \texttt{LAB1}, \texttt{LAB2}, \texttt{CON}, \texttt{EFC}, and \texttt{ACH} exhibit marked variations. In particular for UK 2019, \texttt{LAB2} almost doubled its initial size between November 21 and 26. \texttt{CON} exhibits a similar trend, albeit with reduced magnitude. For USA 2020 instead, \texttt{EFC} initially expanded and subsequently shrank significantly, while \texttt{ACH} lost more than 50\% of its original size in the last couple of weeks before the election. This result demonstrates that some CCs were rather unstable, as they experienced major temporal variations in size. Moreover, our results also demonstrate that some CCs evolved in time in opposite ways. For example, while \texttt{LAB2} and \texttt{CON} increased their size until around November 23, \texttt{LAB1} presents an opposite trend. Similar considerations can be made for \texttt{SNPO}, which increased its size until election day, with respect to \texttt{RCH} that progressively shrank, and for other communities involved in the USA 2020 debate. 

Figures~\mbox{\ref{fig:temporal-stability}}b and~\mbox{\ref{fig:temporal-stability-us}}b provide additional results by tracking how the membership -- that is, the set of users that belong to a community at a given time -- and not just the size, of each CC changes through time. As expected, CCs that experienced major size variations (e.g., \texttt{LAB2}, \texttt{CON}, \texttt{EFC}, \texttt{ACH}) present the largest differences in membership. Interestingly however, also CCs that exhibited moderate size variations (\texttt{LAB1}), or that appeared as overall stable when only considering their size (\texttt{TVT} and \texttt{ASE}), nonetheless feature marked membership differences in time. This result tells us that a stationary size does not necessarily imply stability in terms of members of the community. In fact, some CCs maintained a relatively stable number of members not because of lack of user shifts between CCs, but as a result of a balanced inflow and outflow of members. Figures~\mbox{\ref{fig:temporal-stability}}c and~\mbox{\ref{fig:temporal-stability}}d provide detailed results on this aspect for UK 2019, while Figures~\mbox{\ref{fig:temporal-stability-us}}c and~\mbox{\ref{fig:temporal-stability-us}}d do for USA 2020. As shown, some CCs (\texttt{B60}) had a limited influx and outflux for all the time, which is reflected in Figure~\ref{fig:temporal-stability}b by a relatively stable membership. On the flip side, other CCs (\texttt{RCH}, \texttt{CON}) were much more unstable, with a massive influx and outflux of users that translates into unstable membership. Interestingly, there also exist less straightforward situations, such as those of \texttt{TVT} and \texttt{CRE} that featured strong influx and outflux, but a relatively stable size and membership. This result implies that many users joined and left \texttt{TVT} and \texttt{CRE} every time window, but that those who left were likely to come back at a later time and vice versa. In other words, both \texttt{TVT} and \texttt{CRE} were characterized by a restricted set of users who repeatedly joined and left the communities. Overall, our results provide evidence of temporal instability and highlight marked differences in the temporal dynamics of some CCs.

\subsubsection*{Representativeness of static CCs}
\begin{figure}[!t]
  \centering
  \begin{minipage}{.3\columnwidth}\includegraphics[width=\textwidth]{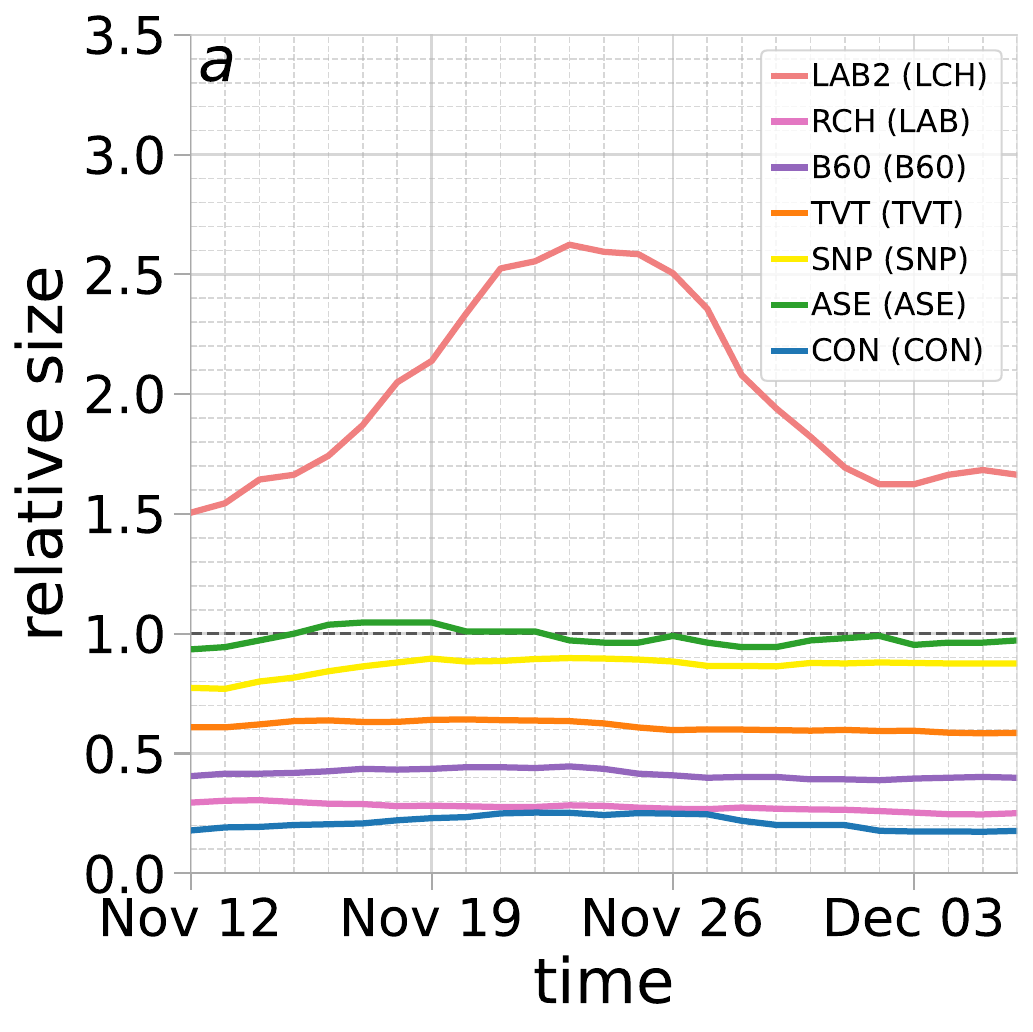}
  \end{minipage}\hspace{.05\columnwidth}\begin{minipage}{.3\columnwidth}\includegraphics[width=\textwidth]{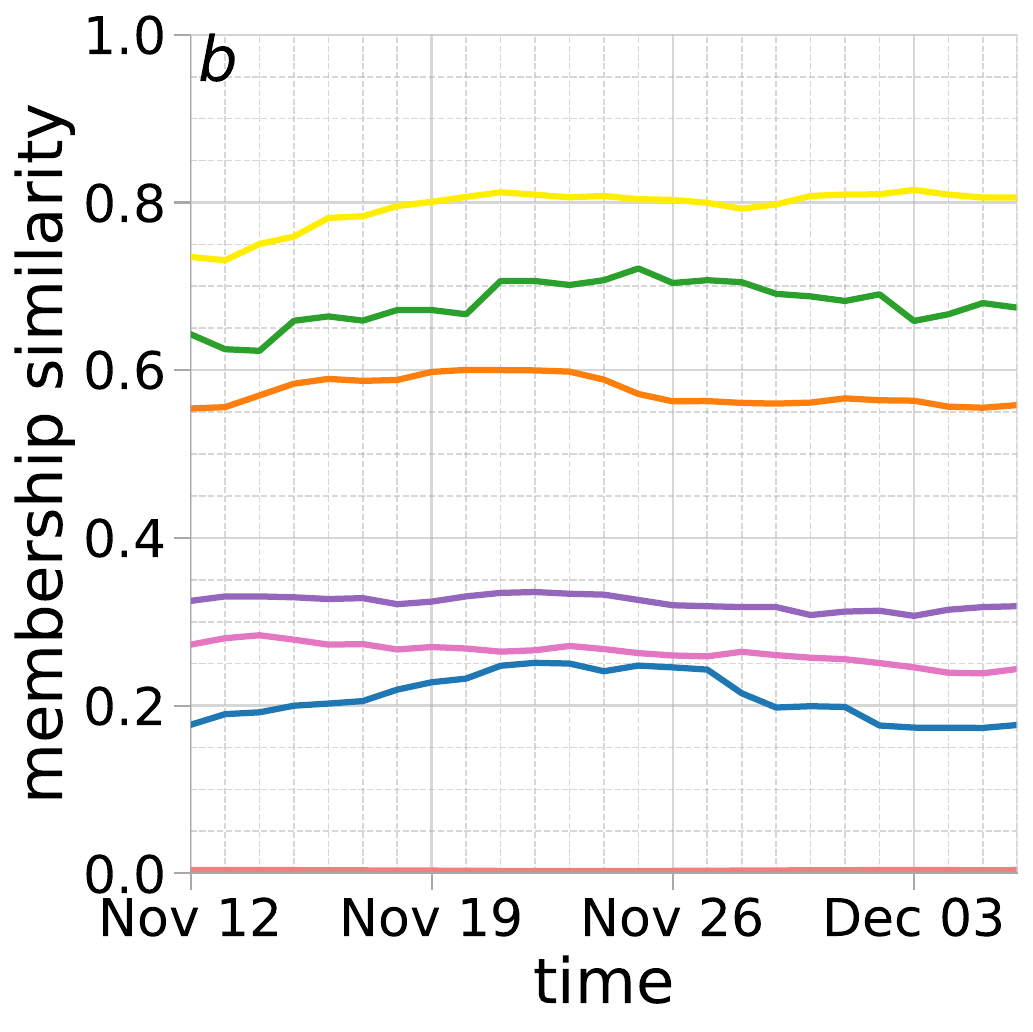}
  \end{minipage}\caption{UK 2019: Comparison between the time-evolving size (a) and membership (b) of each dynamic CC, and the static size and membership of the corresponding static CC.}
  \label{fig:size-jaccard-dynamic-static}
\end{figure}

\begin{figure}[!t]
  \centering
  \begin{minipage}{.3\columnwidth}\includegraphics[width=\textwidth]{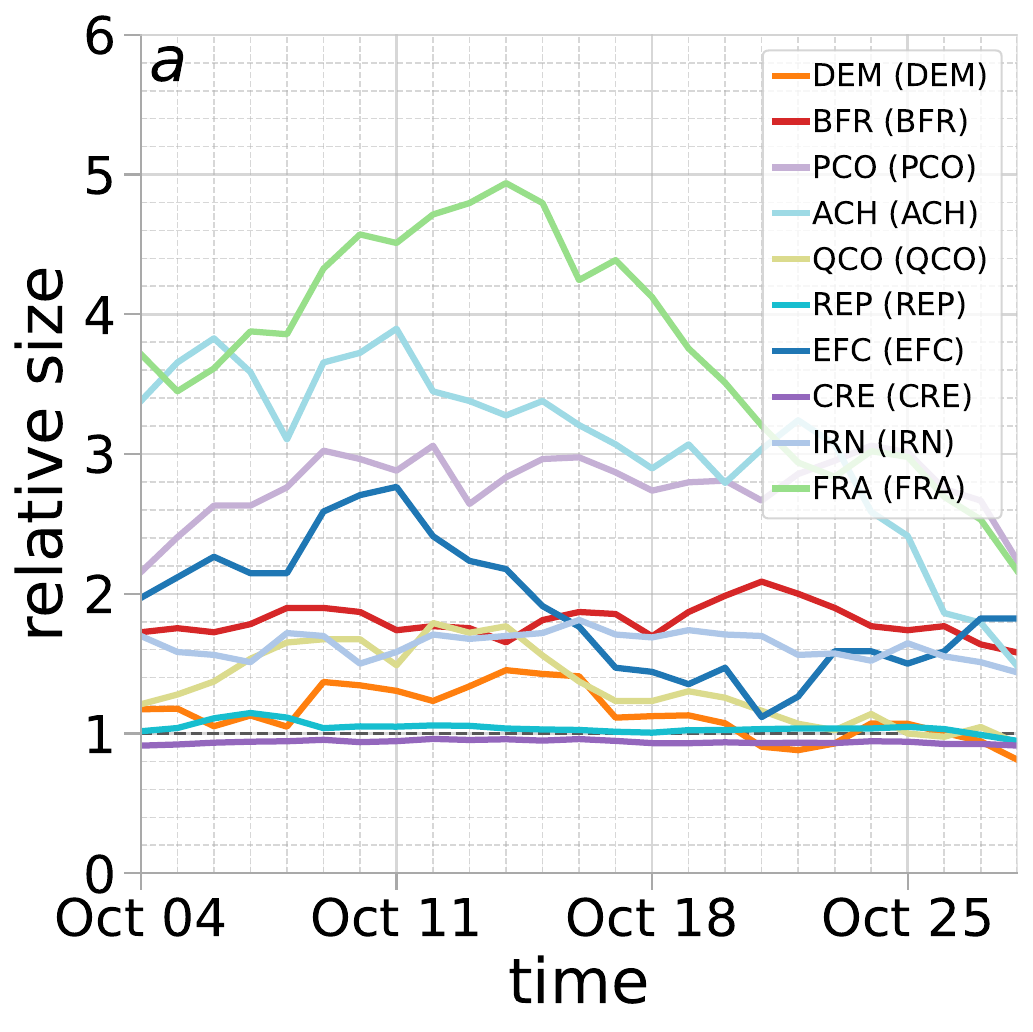}
  \end{minipage}\hspace{.05\columnwidth}\begin{minipage}{.3\columnwidth}\includegraphics[width=\textwidth]{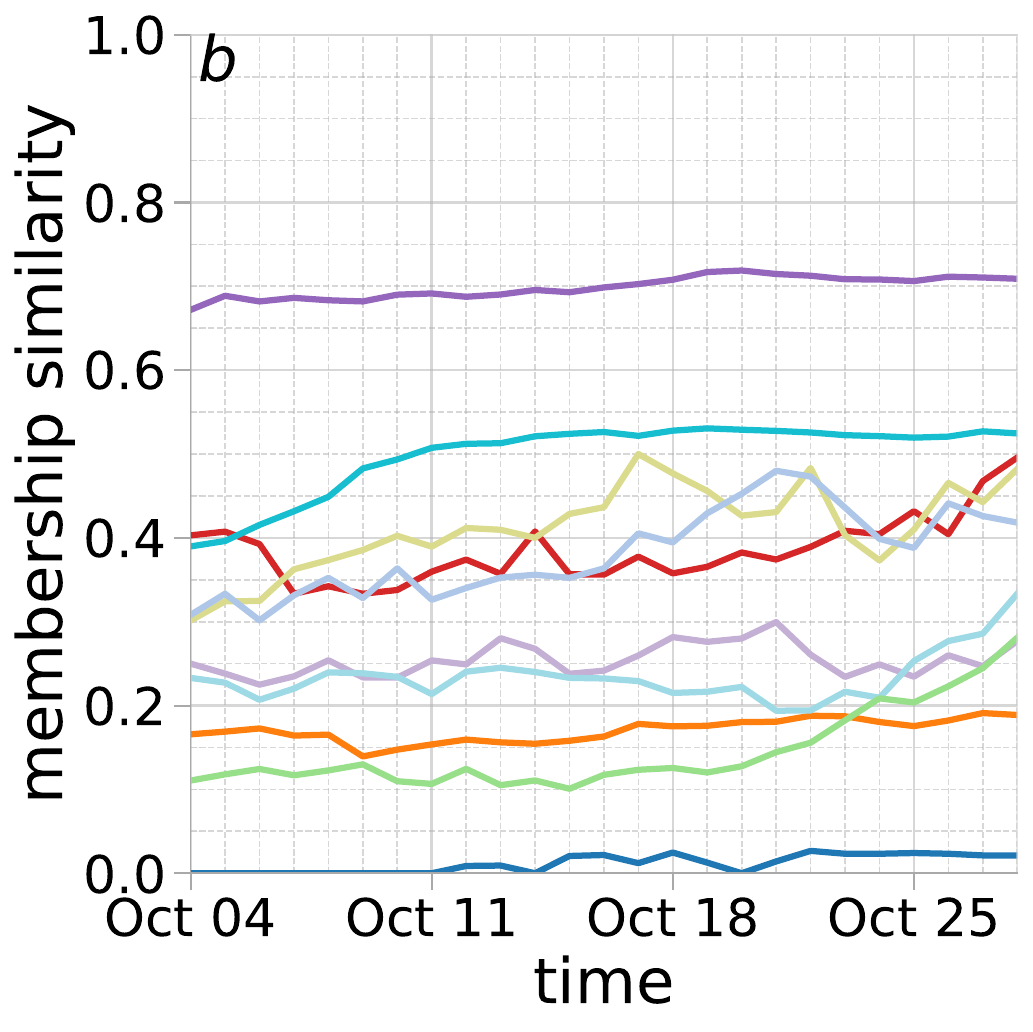}
  \end{minipage}\caption{USA 2020: Comparison between the time-evolving size (a) and membership (b) of each dynamic CC, and the static size and membership of the corresponding static CC.}
  \label{fig:size-jaccard-dynamic-static-us}
\end{figure}

We conclude our analysis of RQ1 by discussing the implications of our findings about the instability of the CCs, with respect to the static analyses of coordinated behavior. In particular, results in both Figure~\mbox{\ref{fig:temporal-stability}} and Figure~\mbox{\ref{fig:temporal-stability-us}} highlighted that some CCs underwent marked changes in time that cannot be captured with time-aggregated static analyses. Figures~\mbox{\ref{fig:size-jaccard-dynamic-static}} and~\mbox{\ref{fig:size-jaccard-dynamic-static-us}} dig deeper into this aspect by comparing the time-evolving size and membership of each dynamic CC to those of its corresponding static CC. In particular, to be strongly representative of a dynamic community, a static CC must have relative size $S$ and membership similarity $J$ both close to 1. The mapping between static and dynamic CCs is in Figure~\ref{fig:static-vs-dynamic}. The results in Figures~\mbox{\ref{fig:size-jaccard-dynamic-static}} and~\mbox{\ref{fig:size-jaccard-dynamic-static-us}}  surface some of the drawbacks of static analyses. The only static CCs that are strongly representative of their dynamic counterparts are \texttt{ASE} and \texttt{SNP} for UK 2019, and \texttt{CRE} for USA 2020, as reflected by relative size $S\simeq1$ and membership similarity $J>0.6$. The same can be concluded also for \texttt{TVT}, \texttt{REP}, and \texttt{QCO}, although to a much lower extent. Instead, all other static CCs are weakly representative of the corresponding dynamic CCs. In fact, all have relative size far from 1 (e.g., $S\simeq5$ on October 14 for \texttt{FRA}), and membership similarity $J<0.4$. Overall, we found several static CCs that are poorly representative of the temporal evolution of the corresponding communities, which could possibly lead to inaccurate or unreliable results.



\subsection*{RQ2: Temporal dynamics of user behavior}
\begin{figure}[!t]
\centering
    \begin{minipage}{.45\columnwidth}\includegraphics[width=\textwidth]{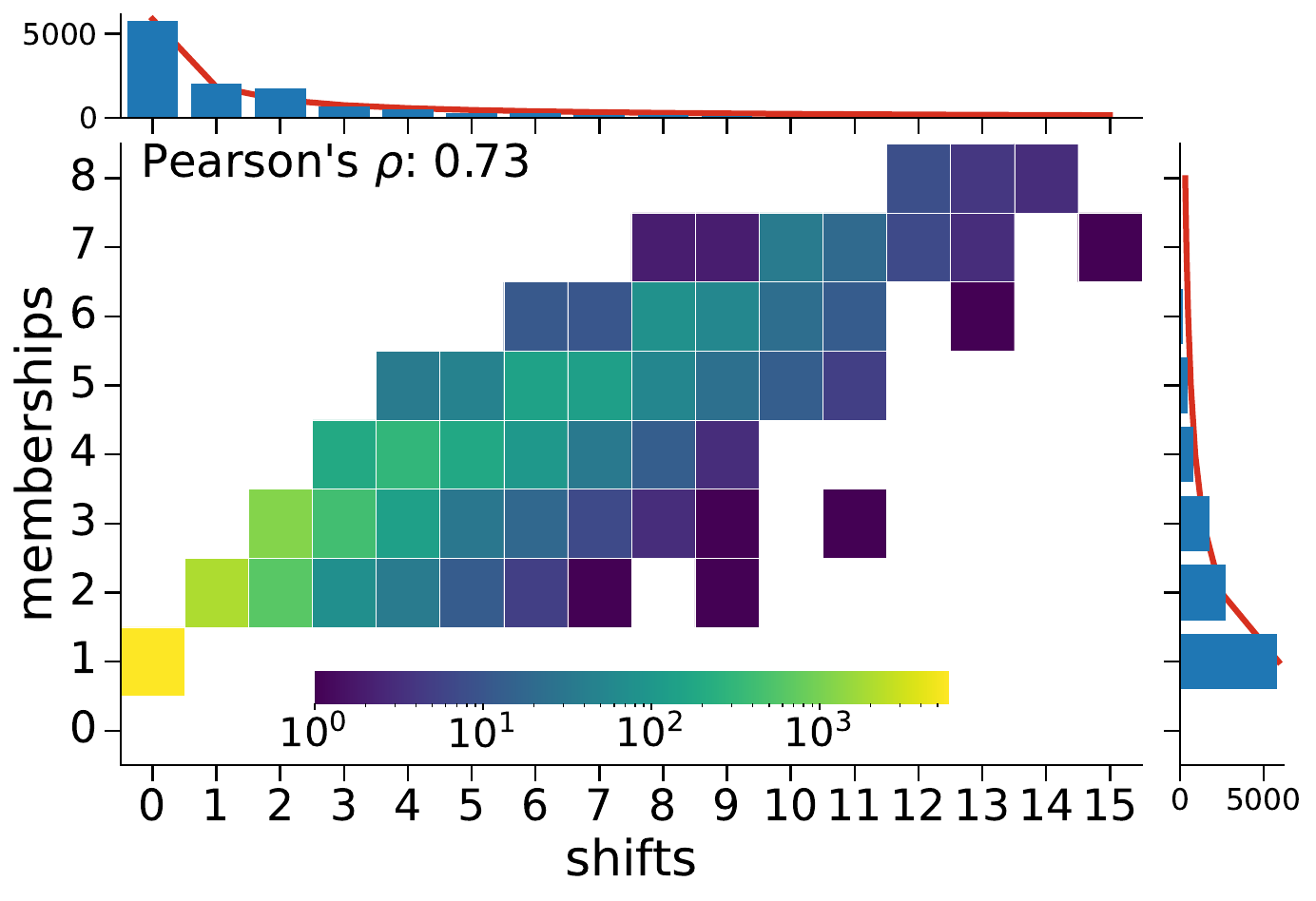}
        \subcaption{UK 2019.}
        \label{fig:user-distributions-uk}
    \end{minipage}\hspace{.05\textwidth}\begin{minipage}{.45\columnwidth}\includegraphics[width=\textwidth]{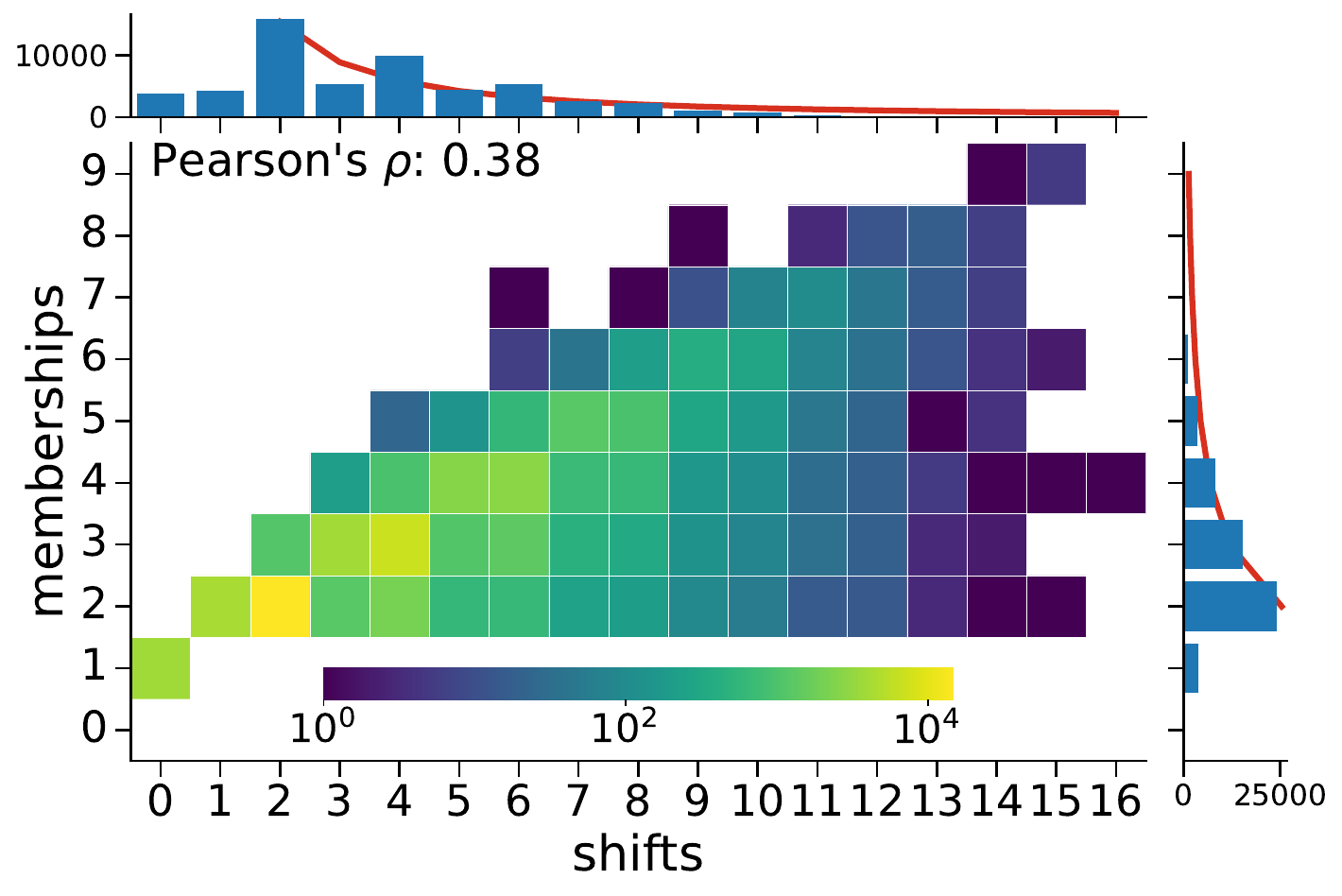}
        \subcaption{USA 2020.}
        \label{fig:user-distributions-us}
    \end{minipage}\caption{Joint discrete distribution of the number of user shifts between CCs and user memberships to CCs, with marginal univariate distributions. Notably: \textit{(i)} 51\% and 93\% of all users belonged to more than one CC for UK 2019 and USA 2020, respectively, \textit{(ii)} memberships and shifts are strongly correlated ($\rho = 0.73$) in UK 2019 and moderately so ($\rho = 0.38$) in USA 2020, \textit{(iii)} the marginal distributions of memberships and shifts are heavy-tailed in both datasets.}
    \label{fig:user-distributions}
\end{figure}

The different temporal evolution of the CCs that we observed in RQ1 are due to different user behaviors, which we investigate in this section. Figure~\ref{fig:user-distributions} shows the joint and marginal distributions of the number of distinct CCs to which users belonged and the number of user shifts between CCs. As shown, $72\%$ of all UK 2019 users and $50\%$ of all USA 2020 users belonged to two or less CCs for all the time, which is also represented by all marginal distributions being skewed towards small numbers of distinct memberships and shifts. In fact, all four marginal distributions in Figure~\mbox{\ref{fig:user-distributions}} are heavy-tailed, as highlighted by the accurate power-law fitting shown with red lines in the marginal histograms. At the same time however, a minority of users belonged to many CCs, which explains our previous results on the instability of some communities. Figure~\mbox{\ref{fig:user-distributions}} also allows evaluating the behavior that we observed in Figure~\mbox{\ref{fig:temporal-stability}} for \texttt{TVT} and in Figure~\mbox{\ref{fig:temporal-stability-us}} for \texttt{CRE}: users repeatedly leaving and re-joining the same community. Such behavior is reflected in Figure~\mbox{\ref{fig:user-distributions}} with users having many shifts but few memberships. Figure~\mbox{\ref{fig:user-distributions-uk}} shows few users with this behavior and, in fact, the distributions of memberships and shifts for UK 2019 are strongly correlated (Pearson's $\rho = 0.73$) meaning that the behavior observed for members of \texttt{TVT} is overall marginal in the whole dataset. On the contrary, Figure~\mbox{\ref{fig:user-distributions-us}} depicts a slightly different situation for USA 2020, as demonstrated by the relatively dense (hot) region of the heatmap below the main diagonal of the plot. This means that the behavior observed for \texttt{CRE} is more prominent, albeit still related to a minority of users  as reflected by the moderate correlation between the distributions of memberships and shifts for USA 2020 (Pearson's $\rho = 0.38$).

The analysis of Figure~\mbox{\ref{fig:user-distributions}} only considers the number of memberships and shifts between CCs. However, not all shifts are the same, as moving between two opposite communities (e.g., at the extremes of the political spectrum) entails a much bigger change -- a farther leap -- than moving between two similar ones. To account for this facet we assign a weight $w_{k,j}$ to all shifts $s_{k\;\rightarrow\;j}$ between any origin community $C_k$ and any destination community $C_j$, based on the (dis)similarity between $C_k$ and $C_j$ (See \textit{Materials and Methods}). When considering also the distance between the origin and destination CCs involved in user shifts, results show that the majority of shifts occur between politically-similar communities. This finding is in line with the theories about political polarization and echo chambers~\mbox{\cite{garimella2018political}}. Nonetheless, a minority of users experienced major ideological shifts by moving across the political spectrum, for both the UK 2019 and the USA 2020 election (See \textit{Supporting Figure}~S3). Another interesting observation derived from this analysis is that the vast majority of shifts for UK 2019 occurred towards the left of the political spectrum. This means that, overall, the users involved in the UK 2019 online electoral debate ideologically moved towards the left as the debate unfolded (See \textit{Supporting Figure}~S3 and corresponding discussion).

\subsection*{RQ3: Archetypes and drivers of user behavior}
When investigating user behaviors in RQ2, Figure~\ref{fig:user-distributions} surfaced heavy-tailed distributions for both user memberships to CCs and shifts between CCs, which imply heterogeneous user behaviors. On one hand these distributions represent a bulk of stationary users with few shifts and memberships. On the other hand however, their long tails also admit the existence of some volatile users characterized by a multitude of shifts, and of all other behaviors in between the ``stationary'' and ``volatile'' extremes. Following this observation, we introduce three archetypes of users, each corresponding to different temporal behaviors with important practical implications. For each archetype, we \textit{(i)} propose an operative definition, \textit{(ii)} apply the definition to measure the presence of such users in our CCs, and \textit{(iii)} explore possible motivations for their behavior.

\subsubsection*{Archetype 1: Stationary}
\begin{figure}[!t]
\centering
    \begin{minipage}{.4\columnwidth}\includegraphics[width=\textwidth]{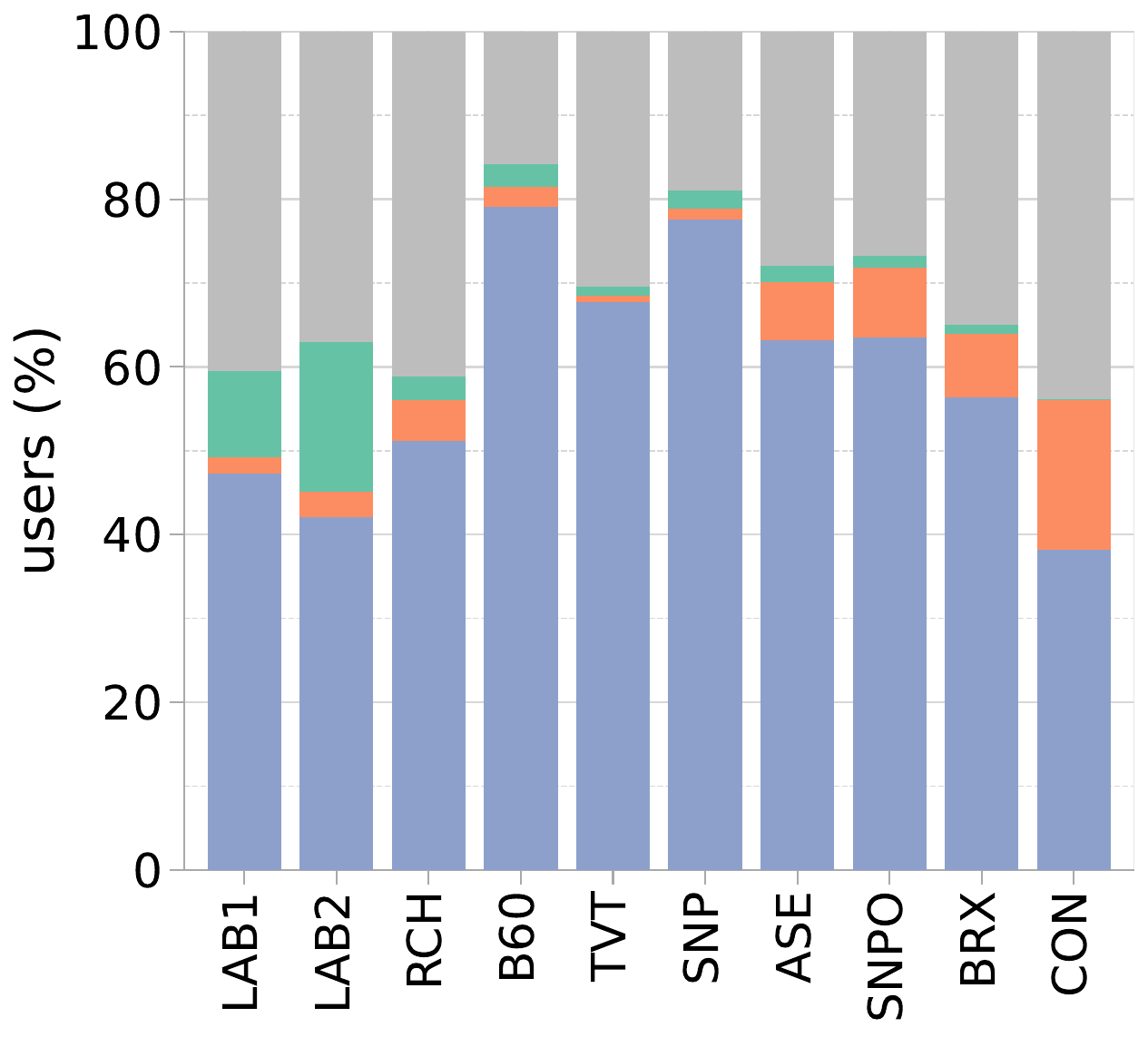}
        \subcaption{UK 2019.}
        \label{fig:archetypes-per-cc-uk}
    \end{minipage}\hspace{.01\columnwidth}\begin{minipage}{.4\columnwidth}\includegraphics[width=\textwidth]{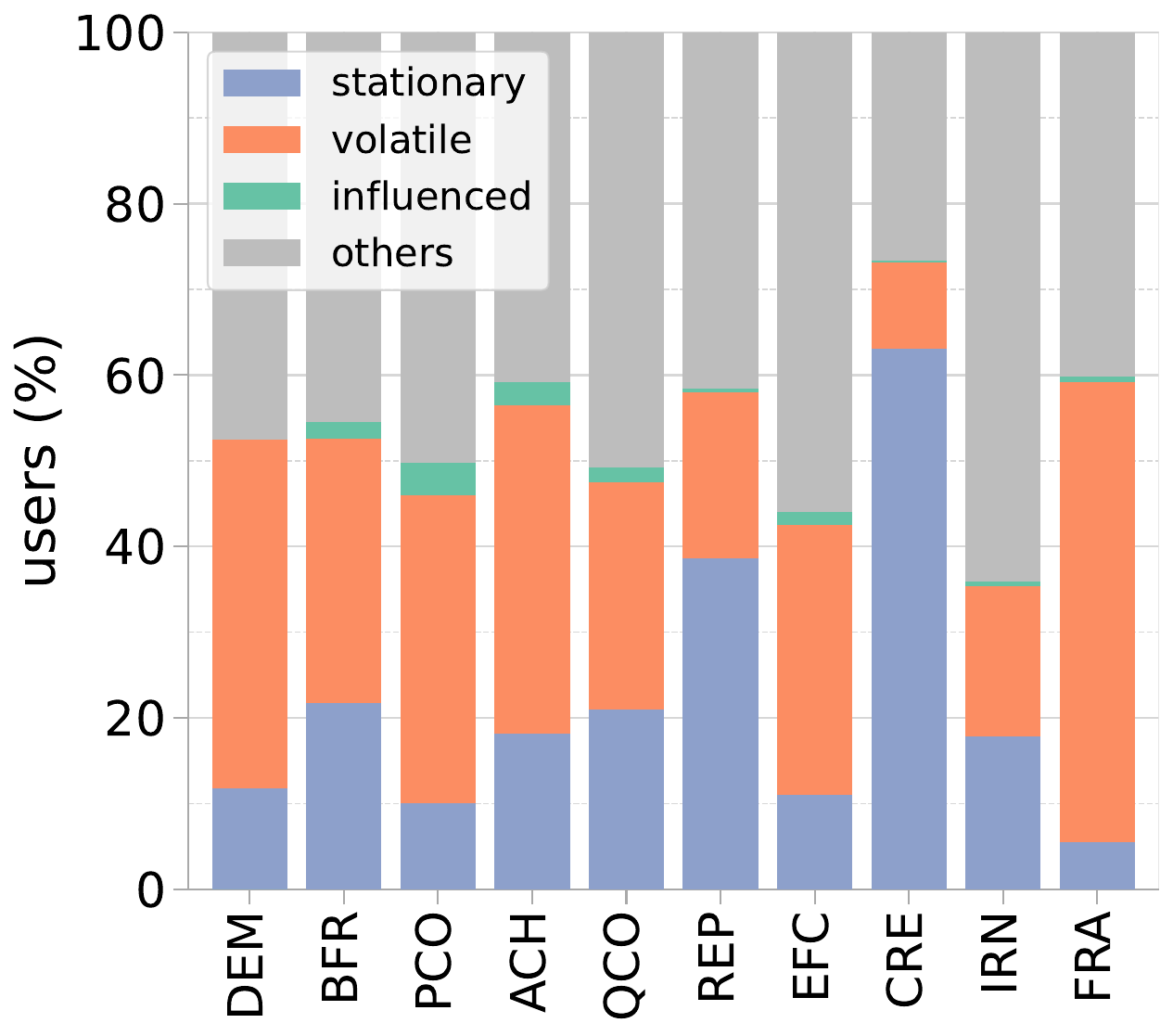}
        \subcaption{USA 2020.}
        \label{fig:archetypes-per-cc-us}
    \end{minipage}\caption{Membership composition of the CCs in terms of the different archetypes of users. CCs are ordered on the \textit{x} axis according to their political leaning. }
    \label{fig:archetypes-per-cc}
\end{figure}

\textit{Stationary users are those who belong to the same community for all the time.} This archetype straightforwardly emerges from Figure~\ref{fig:user-distributions} and the above discussion. The analysis of stationary users is relevant because their behavior could imply that they are strong supporters and core members of their CC~\cite{trujillo2022make}. Figure~\ref{fig:archetypes-per-cc} shows the proportion of stationary users, as well as of the users of the remaining archetypes, in the CCs. Users that do not match the definition of any archetype introduced in this section are grouped as ``others''. As shown, the proportion of stationary users is in the region of 50\% for UK 2019, while it varies significantly for the CCs involved in USA 2020. In detail, center-leaning communities in UK 2019 -- such as \texttt{B60}, \texttt{TVT}, and \texttt{SNP} -- featured a large share of stationary users: between 68\% and 79\% of all community members. Contrarily, stationary users constituted only $\simeq42\%$ of the strongly polarized CCs such as \texttt{LAB1}, \texttt{LAB2}, and \texttt{CON}. This finding suggests that politically polarized communities were more unstable than moderate ones during the UK 2019 online electoral debate. Interestingly, also our results in Figure~\ref{fig:temporal-stability} support this conclusion, with politically extreme CCs appearing as overall more unstable. This result is particularly relevant also in light of the many studies that specifically focus on strongly polarized communities, such as those on political polarization, fringe and extreme behaviors, and far-right online groups~\cite{agathangelou2017understanding,zannettou2018origins,morstatter2018alt}.
The main difference between UK 2019 and USA 2020 is represented by the overall lower fraction of stationary users in the former online debate. Within this context the two Republican CCs were the most stable ones with \texttt{REP} (Republicans) containing 39\% stationary users and \texttt{CRE} (Conspiracist Republicans) containing as much as 62\%. This result is again consistent with those reported in Figures~\mbox{\ref{fig:temporal-stability-us}}a and~\mbox{\ref{fig:temporal-stability-us}}b about the temporal stability of CCs. Stationary users for all other CCs were in the region of 18\%. This marked difference between UK 2019 and USA 2020 is motivated by the different political polarization of the communities that took part in the two online debates, shown in Figure~\mbox{\ref{fig:polarity-spectrum}}. Indeed, while the UK 2019 CCs almost uniformly spanned the whole political spectrum, eight out of ten of those involved in USA 2020 laid on the right-hand side of the spectrum. In turn, this resulted in more shifts between communities and consequently in less stationary users.

To gain insights into why stationary users never leave their CC, we analyze their use of hashtags as a proxy for their interests and viewpoints, in relation to those of their CC and of all other communities (See \textit{Supporting Figure}~S4 and related discussion). The rationale for this analysis stems from the literature on echo chambers and political polarization as these users might be disincentivized to change community because they already belong to the CC that mostly represents their interests and viewpoints~\mbox{\cite{garimella2018political}}. Results show that 94\% of all UK 2019 stationary users and 97\% of all USA 2020 stationary users are mostly similar to the CC to which they belong, as highlighted in figure by the great prevalence of users along the main diagonal. Overall, our results confirm the similarity between the interests and viewpoints of stationary users and those of the CC to which they belong, consistently with the established literature on echo chambers~\mbox{\cite{cinelli2021echo}}.

\subsubsection*{Archetype 2: Influenced}
\textit{Influenced users are those who change community and remain in the destination community for a relatively long time.} One of the advantages of our dynamic analysis lies in the possibility to investigate user shifts between CCs. In the case of polarized or controversial online debates, shifts -- disregarded in previous studies -- provide valuable information on the evolving stance of the users with respect the sides involved in the debate. To this end, users who abandon a community to join another one for a long time might have been persuaded or influenced by the latter. Hence detecting and characterizing influenced users could have important practical implications for the study of online campaigns.
For our subsequent analyses we consider as influenced all those users who never left the destination community $C_j$ after a shift $s_{k\;\rightarrow\;j}$. Moreover, we also impose that each influenced user remained in $C_j$ for at least one third of the time covered in the dataset, so as to avoid considering as influenced those users who stayed in the destination community for just a few days before the election.
Figure~\ref{fig:archetypes-per-cc-uk} shows marked differences in the proportion of influenced users within the UK 2019 CCs. Out of all the communities, \texttt{LAB2} was the one with the highest share (18\%), followed by \texttt{LAB1}, \texttt{B60}, and \texttt{RCH}. These are all left and center-left leaning communities. Opposite results emerge for right-leaning communities, where \texttt{CON} had the lowest share of influenced users, followed by \texttt{BRX}, \texttt{TVT}, and \texttt{SNPO}. These results extend and reinforce those related to the net flows of users between CCs (See \textit{Supporting Figure}~S3). Together, they surface a previously unknown strong political imbalance, in favor of the left, in the capacity to attract and hold users by the CCs involved in the electoral debate. The proportion of influenced users reveals interesting patterns also for the USA 2020 scenario, as shown in Figure~\mbox{\ref{fig:archetypes-per-cc-us}}. Here, the CCs with the highest share of influenced users were \texttt{PCO} and \texttt{ACH}, followed by \texttt{BFR}, \texttt{QCO}, and \texttt{EFC}. In the context of the USA 2020 online debate, these were the most center-leaning communities. Similarly to the UK 2019 case, the analysis of influenced users in USA 2020 unveiled an interesting pattern in the capacity of some CCs to pull and hold users.

Next, we explore possible drivers for the behavior of influenced users. For each influenced user $u$ we investigate whether there exist signals capable of explaining its shift $s_{k\;\rightarrow\;j}$ from the origin community $C_k$ to the destination community $C_j$. For this analysis we consider the time-evolving relationship between $u$, $C_k$, and $C_j$ from a twofold perspective: \textit{(i)} topic-based similarity and \textit{(ii)} topological position in the network. With the former we assess whether influenced users exhibited a time-increasing topic-based similarity with their destination communities. With the latter we evaluate whether the position of the influenced users in the temporal network became gradually closer to their destination community. The temporal trends (See \textit{Supporting Figures}~S5 and~S6, and related discussions) reveal that influenced users indeed exhibited an increasing similarity -- be it topic- or network-based -- to their destination community as time went by. Meanwhile, they also became increasingly dissimilar to their origin community. This finding is confirmed in both datasets and holds under all the viewpoints considered in our analyses, revealing that content and network characteristics provide insights into why users shift between CCs (See \textit{Supporting Information} for details).

\subsubsection*{Archetype 3: Volatile}
\textit{Volatile users are those who repeatedly change community, staying in each community only for a limited amount of time.} At the opposite of stationary users are those whose behavior is very erratic. Volatile users often shift between CCs without attaching to any, if not for a very limited time. The identification and characterization of volatile users in online debates is relevant, as they might represent undecided users who have not already taken a definitive position about the discussed topic~\cite{lenti2022ensemble}. Here, we operationalize volatile users as those who performed three or more shifts, and that spent less than one third of the total time in each CC to which they belonged. Figure~\ref{fig:archetypes-per-cc-uk} shows that the proportion of volatile users in the UK 2019 CCs is skewed towards the right-hand side of the political spectrum. In detail, \texttt{CON} had by far the largest share of volatile users (18\%), followed by \texttt{SNPO}, \texttt{BRX}, and \texttt{ASE} -- which are all right-leaning communities. Volatile users in each of the remaining CCs accounted for less than 5\% of each community's members. This result is the counterpart of what we discussed about influenced users for the left-leaning CCs. In fact, while we previously found evidence of the effectiveness of the left-leaning communities in attracting (i.e., influencing) users as the electoral debate unfolded, here we find evidence of the weakness of the right-leaning CCs. Interestingly, this result appears to be in contrast to the majority of the existing literature on the use of online platforms by political groups, where right-leaning communities were described as tech-savvy and capable of making the most out of social media campaigning~\cite{morstatter2018alt,gonzalez2022advantage}. The distribution of volatile users within the USA 2020 CCs is relatively uniform, with the exception of \texttt{CRE} and \texttt{FRA}. About the former, all the results are consistent in highlighting that CC as the overall more stable. This result is particularly relevant in comparison to the overall instability of the majority of CCs involved in the USA 2020 debate. The latter community instead features an opposite behavior, as it is characterized by a marked instability (Figure~\mbox{\ref{fig:size-jaccard-dynamic-static-us}}a), emphasized by a majority of volatile users and very few stationary ones (Figure~\mbox{\ref{fig:archetypes-per-cc-us}}). All other CCs involved in the USA 2020 debate have a considerable fraction of volatile users in the region of 40\%. Overall, these results indicate that the USA 2020 online debate was characterized by much more instability than the UK 2019 one, as demonstrated by both the results on the fraction of stationary, influenced, and volatile users (Figure~\mbox{\ref{fig:archetypes-per-cc}}) as well as by the analysis of temporal stability of the CCs involved in the two debates (Figure~\mbox{\ref{fig:size-jaccard-dynamic-static}} \textit{vs.} Figure~\mbox{\ref{fig:size-jaccard-dynamic-static-us}}).

When analyzing possible drivers for the behavior of volatile users, we found that the shifts by volatile users are typically very short (See \textit{Supporting Figure}~S7). This result is again consistent with the echo chamber theory~\mbox{\cite{cinelli2021echo}} and explains why volatile users change community so often. Interestingly, the same analysis also shows that some influenced users permanently join politically-distant communities.

\begin{figure}[!t]
\centering
    \includegraphics[width=.85\columnwidth]{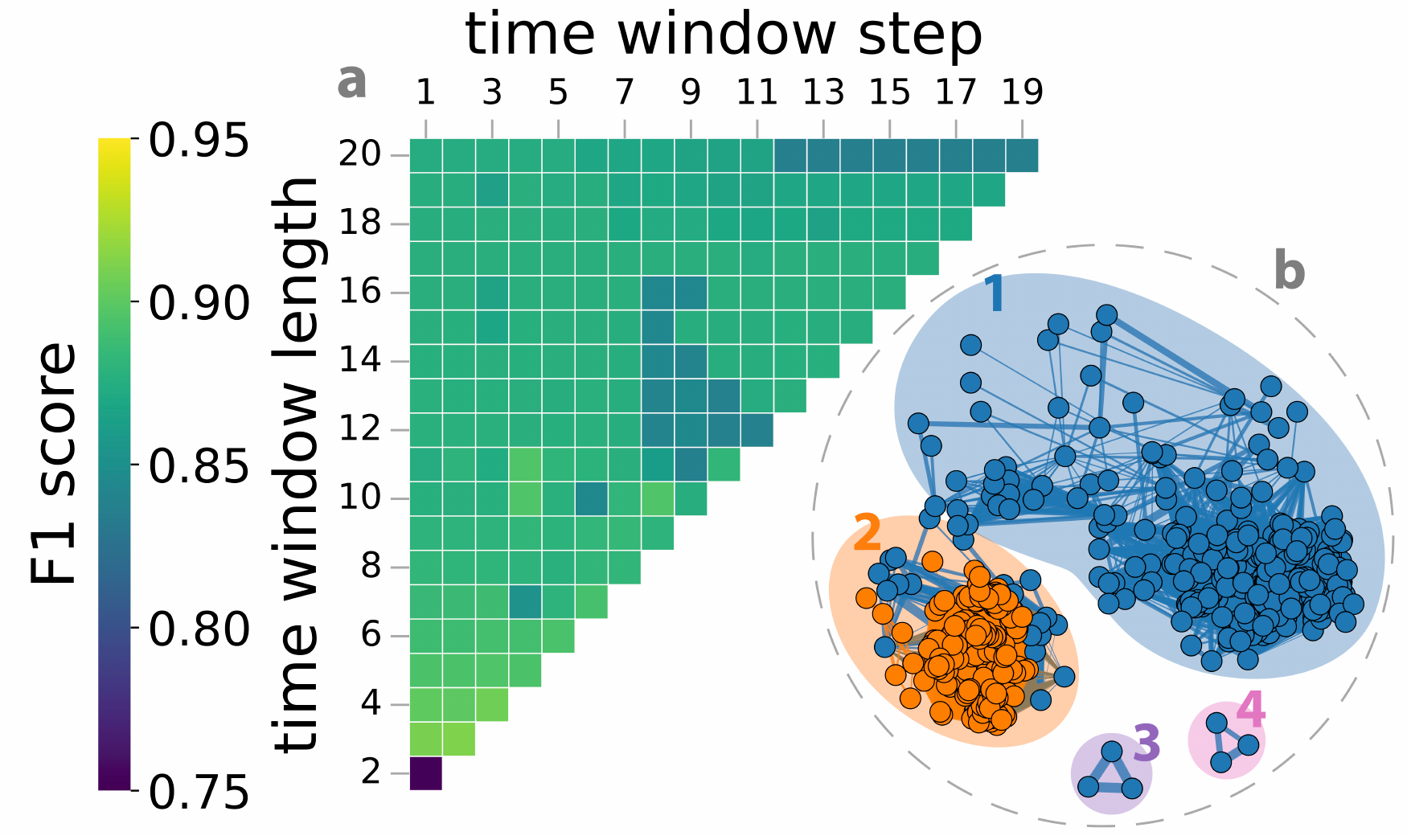}
    \caption{Honduras 2019. (a) Efficacy (F1 score) of our method at detecting coordinated communities involved in an information operation, based on the choice of parameters \textit{time window length} and \textit{time window step}. The resolution parameter $\gamma$ of the community detection algorithm is set to the optimal value $\gamma = 10^{-5}$. (b) Snapshot (layer) of the Honduras 2019 multiplex temporal network showing that all inauthentic accounts (orange-colored) are clustered in a single CC, together with only a few genuine accounts (blue-colored).} \label{fig:grid-heatmap}
\end{figure}

\subsection*{Validation}
As a final case study, we apply our method to a Twitter dataset related to a large-scale information operation (See \textit{Materials and Methods} and \textit{Supporting Information} for details on the dataset). The dataset is composed of the inauthentic accounts that took part in the operation and of a comparable set of genuine accounts. As such, it represents a labeled benchmark suitable for validating the efficacy of our method at distinguishing between inauthentic and inorganic forms of coordination versus genuine ones. Specifically, we run our method with different combinations of parameters, evaluating at each run the extent to which the inauthentic accounts are grouped in distinct communities with respect to the genuine ones. To measure the extent of separation between inauthentic and genuine accounts, we adapt the well-known F1 score to our problem (See \textit{Materials and Methods}). Figure{~\ref{fig:grid-heatmap}}a shows an excerpt of the results that we obtained with a grid search over the two main parameters of our method (i.e., the \textit{length} and \textit{step} of the time windows), when fixing the resolution parameter of the community detection algorithm. As shown, the application of our method yields coordinated communities in which the inauthentic accounts are generally separated from the genuine ones, as testified by $0.75 \le \text{F1} \le 0.91$. The analysis reveals a trend where larger F1 scores are obtained for small values of the two parameters. In fact, the best F1 score is obtained with time window length = 3 days and step = 2 days. Figure{~\ref{fig:grid-heatmap}}b shows one layer of the multiplex temporal network where the nodes are colored so as to represent the inauthentic (orange-colored) and genuine (blue-colored) accounts, and where the different coordinated communities are highlighted. The network in figure shows a good separation between the inauthentic and genuine accounts, where the former are all grouped in a single community with only a few genuine nodes. For reference, applying the traditional static coordination detection approach to this problem would result in F1 $= 0.82$, demonstrating the advantage of a properly configured dynamic approach.  \section*{Discussion}
\label{sec:discussion}
We investigated the temporal dynamics of coordinated online behavior in the context of two recent major elections and an information operation. Our novel approach, grounded on a multiplex temporal network and dynamic community detection, identified more coordinated communities than those found in previous works, surfacing temporal nuances of coordination that would not be observable with the traditional static approach. Our analysis produced key findings in multiple areas. Regarding coordinated communities, we found that the majority were rather unstable and experienced significant changes in size and membership through time. Regarding user membership to coordinated communities, we found evidence of heavy-tailed distributions implying the existence of a bulk of relatively stationary users and of a long tail of volatile ones. The majority of user shifts from a community to another occurred between politically-aligned communities according to the echo chamber theory, although we found a subset of users who crossed the political spectrum. Finally, we found that content and network characteristics convey useful information for understanding why users move between certain coordinated communities.

\subsection*{Implications} Our results bear important implications about the limitations of static analyses of online coordination, the strategies of coordinated communities, the patterns of online influence, and for the research and policy of online platforms.

\subsubsection*{Instability and influence}
The first major implication of our study is the increased awareness of the instability of coordinated communities. To this end, our results open up new directions of research on the temporal dynamics of coordinated online behavior, which was so far almost exclusively analyzed from a static rather than a \textit{dynamic} standpoint~\mbox{\cite{pacheco2020uncovering,cinelli2022coordinated}}. However, our results show the limitations of the former approach, suggesting that additional efforts should be devoted to devising nuanced and reliable dynamic analyses. Then, our results about the broad array of diverse user behaviors, and the resulting user archetypes that we identified, have significant implications for the study of online interactions and influence~\mbox{\cite{rathje2021out}}. For example, future computational and social research should extensively investigate the motivations for such heterogeneous behaviors, for which we provided but some initial explanations. Our results in this area can be particularly useful towards the ongoing research on online influence and persuasion, as the shifts detected via dynamic analyses of coordination could contribute to identifying successful cases of influence over users or communities in a network~\mbox{\cite{smith2021automatic}}. Similarly, our scientific approach could be carried over to investigate the temporal dynamics of online polarization and the possible temporal evolution of echo chambers~\mbox{\cite{cinelli2021echo,tokita2021polarized}}. Our findings are also particularly impactful towards understanding, and possibly even predicting, the outcome of polarized and controversial online debates, such as those preceding major elections~\mbox{\cite{metaxas2012social}}. Specifically, if verified in other contexts and platforms, our results could provide useful information for nowcasting and forecasting the dynamics of group influence in online debates. 

\subsubsection*{From observational to predictive studies}
In the future it would be possible to progressively shift from observational to \textit{predictive} analyses, provided the availability of adequate ground-truths, which currently represents a limiting factor for all studies in the area of coordinated online behavior~\mbox{\cite{vargas2020detection}}. In addition to opening up the possibility to predict the dynamics of online influence, our results show interesting temporal correlations between significant changes in the structure of coordinated communities and real-world events (See \textit{Supporting Figure}~S8 and related discussion). Therefore, while additional research is needed to design the methods and validate the approach, our results open the door to: \textit{(i)} leveraging external knowledge about real-world events to estimate their online impact in terms of the changes experienced by the interested communities; \textit{(ii)} leveraging significant changes in community structure to identify events that have been noteworthy in a specific period of time and for a specific audience~\mbox{\cite{kalyanam2016prediction}}.

\subsection*{Validation}
The lack of reference datasets and authoritative ground-truths on coordinated online behavior currently represents one of the strongest limiting factors to the research in this area. In addition to carrying out manual investigations (See \textit{Supporting Figure}~S8 and the related analysis), here we circumvented this problem by validating our approach on a labeled dataset that contains both inauthentic accounts involved in an information operation and genuine ones. This is a favorable approach to validating coordination detection methods for multiple reasons. First, it provides a way to test the efficacy of a method at capturing the behavioral patterns of different accounts, including the inauthentic ones that should stand out from the rest. To this end, our results revealed that our method allows distinguishing inorganic coordinated communities from organic ones, to a large extent. Second, the same analysis can also inform the choice of parameters of the method, which represents a further outstanding challenge in the field. Finally, it demonstrates the practical usefulness of the dynamic analysis of coordinated behavior for a relevant computational social science task.

\subsection*{Profiling}
While we provided multiple findings towards improving our understanding of online debates and online human dynamics at large~\mbox{\cite{cresci2020emergent}}, we only scratched the surface of a complex and multifaceted phenomenon. Another aspect worthy of discussion is the potential for using our proposed methodology to study and contrast online information manipulation or other nefarious instances of online coordination~\mbox{\cite{lazer2018science}}. \textit{Profiling coordination} -- that is, inferring the main characteristics, peculiarities, and possibly even the intent behind different groups of coordinated users, turns out to be particularly challenging~\mbox{\cite{vargas2020detection}}. However, a certain degree of success at profiling coordination is needed for being able to tell the difference between inauthentic or harmful coordination and unintentional coordination among independent users (e.g., fandoms or other grassroots movements). Here, we intentionally kept a neutral stance with respect to the many existing forms of online coordination, enabling the study of diverse instances of coordination without any inherent bias. This was because our main goal for this study was to investigate the temporal dynamics of coordination, leaving the task of distinguishing between inauthentic/harmful and authentic/harmless coordination for subsequent analyses. These would ideally be conducted by human analysts, possibly domain experts, who could draw upon the insights and characteristics of the detected communities, as provided by this and other studies~\mbox{\cite{pacheco2020uncovering,nizzoli2021coordinated}}. Nevertheless, we acknowledge the importance of investigating inauthentic and harmful communities as part of future dynamic analyses on online coordination. For example, many harmful instances of coordination, such as strategic information operations, exploit multiple tactics and resources to boost their chances of success~\mbox{\cite{starbird2019disinformation}}. Studying online debates by means of dynamic analyses could allow spotting early signals of harmful coordination, possibly enhancing timely responses to what threatens the safety and integrity of the online environment~\mbox{\cite{bak2021stewardship}}.

\subsection*{Limitations and Future Work}
The main drawback of our study stems from its relatively limited scope, with respect to the breadth and depth of a complex phenomenon such as online coordination. For example, we do not make any assumption on the possible inauthenticity or harmfulness of the identified coordinated communities of UK 2019 and USA 2020. However, inauthentic and harmful coordination could be used in the context of online political debates as a mean to influence the electoral outcome~\mbox{\cite{bail2020assessing}}. Here, our results for the USA 2020 presidential election surfaced the activity of multiple coordinated groups supporting election fraud narratives and other conspiracy theories~\mbox{\cite{vishnuprasad2024tracking}}. Likewise, we also investigated the behavior of foreign influence groups that participated in the electoral debate. Nevertheless, our results do not specifically address the impact that inauthentic and harmful groups could have had on the analyzed online debates, for which our study does not provide conclusive results, but rather calls for additional research. Moreover, our validation of the methodology, while showing limited sensitivity to small variations in the time window length and step parameters, revealed our method's sensitivity to the resolution parameter of the underlying community detection algorithm (See \textit{Supporting
Information}). However, this is a widely recognized phenomenon in the literature~\mbox{\cite{dao2020community,fortunato2007resolution}}. In fact, the optimal value for the resolution parameter varies based on network characteristics and the goal of the analysis, leaving its selection to the analyst discretion. Finally, our analysis is based on data collected from a single platform, while online coordination often involves activities intertwined across multiple platforms~\mbox{\cite{ng2022coordinating}}. As such, we might have missed significant coordinated efforts that have occurred on platforms other than Twitter. This limitation is shared with the vast majority of the existing literature on the subject, mainly due to the challenges of acquiring related and comparable datasets across multiple platforms. In light of this widespread limitation, however, future research on coordinated online behavior should strive to collect and analyze multiplatform datasets, for that could reveal patterns of coordination that would otherwise remain hidden. Unfortunately however, the limitations related to data availability extend beyond the challenges of studying multiplatform coordination. The recent changes in API availability enforced by Twitter\mbox{~\cite{academics}}, particularly after its transition to X.com, represent a paramount example. The discontinuation of Twitter Academic APIs and the prohibitive costs of all other options pose significant obstacles to data availability, with nefarious consequences in terms of greatly reduced platform transparency and reproducibility of scientific results. While the general outlook on social media data availability remains grim at the time of writing, the European Digital Services Act (DSA) could turn the tide in the struggle for access to platform data, by providing legal and technical facilities for submitting data access requests for research purposes~\mbox{\cite{eu2020DSA}}. Finally, future research should also deviate from the traditional focus of online coordination studies (e.g., political discussions and inauthentic coordination) to encompass topics, temporal dynamics, and coordination patterns that are typical of non-polarized or non-controversial online interactions.
 \section*{Materials and Methods}
\subsection*{Data} The data for our study covers two recent major political events -- the 2019 UK general elections and the 2020 USA presidential elections -- and a large-scale information operation.

\subsubsection*{UK 2019 General Election} We leverage a publicly available reference dataset related to the online Twitter debate about the 2019 UK general election.\footnote{\url{http://doi.org/10.5281/zenodo.4647893}} This dataset is relevant for our present study since it has already been the subject of static analyses of coordinated behavior~\cite{nizzoli2021coordinated,hristakieva2022spread}. The dataset was built in~\cite{nizzoli2021coordinated} via the Twitter Streaming API during the last month before the UK 2019 election day, namely between November 12 and December 12, 2019. It contains 11,264,820 distinct tweets (left- and right-leaning, as well as neutral) about the 2019 UK general election published by 1,179,659 distinct users. The tweets were selected according to a set of hashtags or relationships to party accounts (See \textit{Supporting Information} for details).

\subsubsection*{USA 2020 Presidential Election} To assess the generalizability of our findings we collected and publicly shared a second dataset of tweets related to the USA 2020 presidential election.\footnote{\url{http://doi.org/10.5281/zenodo.7358386}} Similarly to the UK 2019 dataset, this data collection spanned the month leading up to the election day, encompassing the period from October 4 to November 3, 2020. Likewise, this collection process relied on a combination of election-related hashtags, party hashtags, and official party and leader accounts (See \textit{Supporting Information} for details). In total, this dataset comprises 263,518,037 distinct tweets that pertain to the online discourse surrounding the election. The tweets were generated by 15,288,527 distinct users.

\subsubsection*{Honduras 2019 Information Operation} To validate our method we use a publicly available dataset related to a large-scale information operation promoted by the government of Honduras in 2019.\footnote{\url{https://doi.org/10.5281/zenodo.10650967}} The dataset includes both the inauthentic accounts who took part in the operation, as detected by Twitter Moderation Research Consortium,\footnote{\url{https://transparency.twitter.com/en/reports/moderation-research.html}} and a comparable set of legitimate accounts (See \textit{Supporting Information} for details). As such, it can be used as a ground-truth and benchmark to test the efficacy of our method at detecting the coordinated inauthentic accounts. Similarly to the previous cases, we used one month of data, spanning from November 11 to December 11, 2019. This portion of the dataset contains 251,191 tweets shared by 75,845 distinct users.

\subsection*{Dynamic analysis of coordinated online behavior} To detect and study coordinated behavior we broadly follow the state-of-the-art network analysis frameworks recently proposed by~\cite{nizzoli2021coordinated,pacheco2020uncovering,weber2021amplifying}, which consider repeated similarities in user behaviors as a proxy for coordination. However, we differentiate from the existing methods by building a dynamic multiplex network instead of a static one, and by analyzing it with a dynamic community detection algorithm. Our detailed methodology is presented in the following.

\subsubsection*{Preliminaries} To enable comparisons with previous works we compute user similarity based on \textit{co-retweets} -- the action of retweeting the same tweet by different users, and we analyze \textit{superspreaders} -- the top 1\% users with the most retweets.\footnote{Despite accounting for only 1\% of all users, superspreaders produced $\sim$39\% of all tweets and $\sim$44\% of all retweets.} We computed user similarities based on co-retweets also because our study specifically focuses on superspreaders, who are characterized by a marked retweeting behavior~\mbox{\cite{pei2014searching}}, and because of the excellent results obtained with co-retweets in recent related literature~\mbox{\cite{pacheco2020uncovering,schoch2022coordination,weber2021amplifying}} and in our previous static analyses of online coordination~\mbox{\cite{nizzoli2021coordinated}}.

\begin{figure}[!t]
    \centering
    \includegraphics[width=0.5\columnwidth]{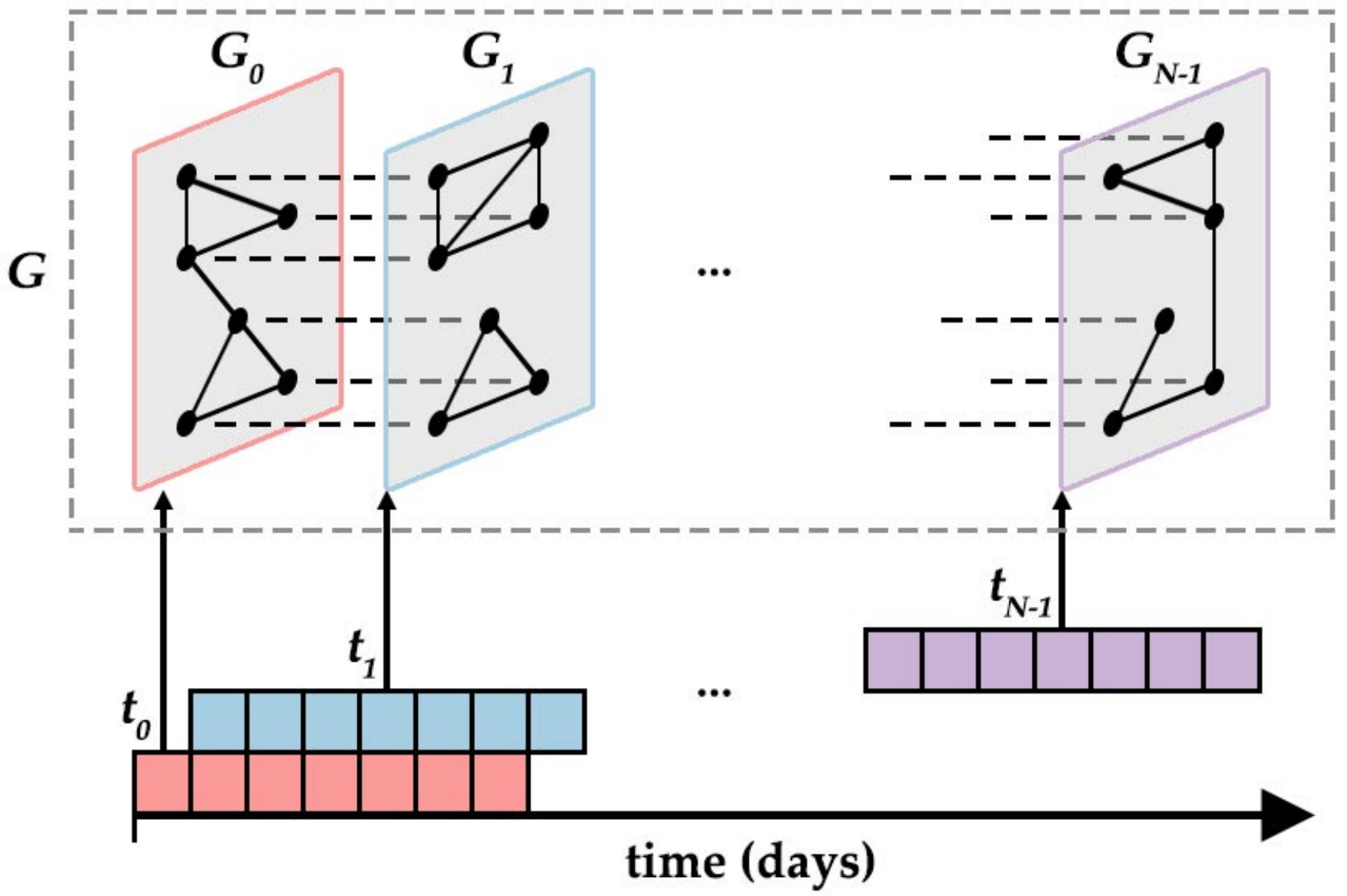}
    \caption{Overview of our methodological approach for building the multiplex temporal network $G$. Data from the overlapping time windows $t_i$ yield the weighted undirected user similarity networks $G_i$ that constitute the layers of $G$. Then, dynamic community detection is performed on the multiplex temporal network $G$.}
    \label{fig:multilayer-net}
\end{figure}

\subsubsection*{Dynamic network modeling} We build our dynamic user similarity network as a multiplex temporal network $G = \langle G_0, \ldots, G_{N-1} \rangle$ where each layer $G_i$ models user behaviors occurred during a given time window $t_i$. We adopted this network representation over other options, such as discrete-time dynamic graphs and embeddings methods, since multiplex temporal networks are state-of-the-art for observational and descriptive tasks~\mbox{\cite{rossetti2018community}}. For example, this representation makes detecting and interpreting perturbations of the network topology (i.e., investigating the stability of CCs) advantageous. We work with a sequence of discrete and overlapping time windows: each $t_i$ has a duration $d=7$ days and an offset (step) $\delta=1$ day from $t_{i-1}$. After a set of preliminary tests we ended up using overlapping time windows instead of non-overlapping ones, because the latter neglect all interactions occurring across any two adjacent windows. On the contrary, overlapping time windows allow to consider all relevant interactions, while also guaranteeing smooth transitions between different time steps. Regarding the size of the time window, the larger it is, the smoother are the transitions between the different snapshots of the multiplex temporal network. However, very large windows hide changes that occur within that time frame, up to the point that particularly large ones make the dynamic analysis collapse to a static one, negating the advantages of the former. On the contrary, particularly small windows may fail to collect meaningful statistics at each time step. Here, our choice of time window length and step is supported by the favorable results obtained for that parameter configuration in our grid search validation. 
As sketched in Figure~\ref{fig:multilayer-net}, for each $t_i$ we build a weighted undirected user similarity network $G_i = (V, E, W)$. To obtain $G_i$ we first model each user $v \in V$ with the TF-IDF weighted vector of the tweets it retweeted during $t_i$. An edge $e \in E$ between two users exists if they retweeted at least one common tweet during $t_i$. Edge weights $w \in W$ are computed as the cosine similarity between pairs of user vectors. The TF-IDF weighting scheme discounts retweets of viral tweets and emphasizes user similarities due to unpopular tweets, contributing to highlighting interesting behaviors. Finally, for each network $G_i$ we retain only the statistically significant edges by computing its multiscale backbone~\cite{serrano2009extracting}.

\subsubsection*{Dynamic community detection} The multiplex temporal network $G = \langle G_0, \ldots, G_{N-1} \rangle$ is suitable for being analyzed with a dynamic community detection algorithm. Leiden is a state-of-the-art community detection algorithm for multiplex networks that improves the well-known Louvain algorithm by identifying higher quality and well-connected communities~\cite{traag2019louvain}. It allows community detection on multiplex networks by jointly considering the internal edges in each layer (solid edges in Figure~\ref{fig:multilayer-net}), as well as the edges that connect nodes across layers (dashed edges in Figure~\ref{fig:multilayer-net}). Leiden is therefore a cross-time algorithm that identifies communities based on the full temporal evolution of the network~\cite{rossetti2018community}. An important implication is the possibility to assign nodes to different communities depending on the time window. As such, it is particularly suitable for studying the temporal evolution of user behaviors and of the CCs. Notably, it would have been possible to use a static community detection algorithm such as Louvain on each snapshot (i.e., layer) of our multiplex temporal network. However, such an approach would have later required the application of a community tracking algorithm, which would have made the overall process more convoluted and error-prone.

\subsection*{Political polarization} We compute a polarization score for each CC based on the political polarization of the hashtags used by its members. We obtain a polarization score for each hashtag in the dataset by applying a label propagation algorithm. In detail, the score for any given hashtag is iteratively inferred from its co-occurrences with other hashtags of known polarity. We initialize the algorithm with the hashtags used for collecting the tweets as the seeds of known polarity (See \textit{Supporting Tables}~S1 and~S2). Once hashtag polarities are computed, the polarity of a CC is obtained as the TF weighted average of the polarities of the hashtags used by members of that community. As a result, each CC is assigned a polarity score $p \in [-1,+1]$. Finally, the scores are normalized 
so that the rightmost-leaning CC in each dataset has $p=+1$ and the leftmost-leaning one has $p=-1$. The most polarized communities are \texttt{CON} ($p=+1$) and \texttt{LAB1} ($p=-1$) for UK 2019, and \texttt{IRN} ($p=+1$) and \texttt{DEM} ($p=-1$) for USA 2020.

\subsection*{Temporal community monitoring} To assess how CCs changed through time we measure, for each community $C_k$, and for each $t_i$ with $i = 0, \ldots, N-1$:
\begin{itemize}
    \item the size of $C_k$ relative to $t_0$: $S(k,i) = \frac{\text{size}_k(t_i)}{\text{size}_k(t_0)}$;
    \item the Jaccard similarity of the membership of $C_k$ relative to $t_0$: $J(k,i) = \frac{|\text{mem}_k(t_i) \; \cap \; \text{mem}_k(t_0)|}{|\text{mem}_k(t_i) \; \cup \; \text{mem}_k(t_0)|}$, where $\text{mem}_k(t_i)$ is the set of users that belong to $C_k$ at time $t_i$;
    \item the influx $F_\text{in}(k,i)$ and outflux $F_\text{out}(k,i)$ of $C_k$, respectively expressed as the cumulative number of users who joined and left the community up to $t_i$.
\end{itemize}

\subsection*{Similarities between communities} We assign a weight $w_{k,j}$ to all shifts $s_{k\;\rightarrow\;j}$ between any origin community $C_k$ and any destination community $C_j$, based on the (dis)similarity between $C_k$ and $C_j$. We compute the similarity between two CCs as the cosine similarity of the TF weighted vectors of the hashtags used by the communities. Then, we weight shifts proportionally to the dissimilarity of the involved CCs: $w_{k,j} = 1 - \text{sim}(k,j)$. We also compute \textit{net shifts} between CCs, as in \textit{Supporting Figure}~S3 where an edge $C_k \rightarrow C_j$ exists only if there is a positive net user flow $F_{k\;\rightarrow\;j}$ from $C_k$ to $C_j$: $F_{k\;\rightarrow\;j} = \sum{s_{k\;\rightarrow\;j}} - \sum{s_{j\;\rightarrow\;k}} > 0$. Then, edge thickness in figure is proportional to $w_{k,j} \times F_{k\;\rightarrow\;j}$. 

\subsection*{Detecting accounts involved in an information operation} We measure the extent to which our method is capable of detecting the inauthentic accounts who took part in an information operation by adapting the well-known F1 score to our problem. For each layer $G_n$ of the multiplex temporal network $G$, we first discard all communities that do not contain any inauthentic account. For each remaining community $C_i^n \; \text{with} \; i=1,\ldots,M$, we compute the Precision $P_i^n$, Recall $R_i^n$, and F1 score $\text{F1}_i^n = 2\frac{P_i^n \times R_i^n}{P_i^n + R_i^n}$. F1 score is the harmonic mean of Precision and Recall, and measures the extent to which the community $C_i^n$ contains all and only the inauthentic accounts. The F1 score for $G_n$ is the weighted mean of the F1 scores of its communities: $\text{F1}_n = \frac{1}{M}\sum_{i=1}^M w_i \text{F1}_i^n$, where $w_i$ is proportional to the fraction of users in $C_i^n$ over all users in $G_n$, so that larger communities contribute more towards $\text{F1}_n$. Finally, the overall F1 score is computed as the mean of the $\text{F1}_n$ scores of each layer $G_n \in G$. \section*{Acknowledgements}
This work was partially supported by project SERICS (PE00000014) under the NRRP MUR program funded by the EU -- NGEU and by SoBigData.it which receives funding from European Union – NextGenerationEU – National Recovery and Resilience Plan (Piano Nazionale di Ripresa e Resilienza, PNRR) – Project: “SoBigData.it – Strengthening the Italian RI for Social Mining and Big Data Analytics” – Prot. IR0000013 – Avviso n. 3264 del 28/12/2021. S.C. is supported in part by the European Commission (ERC-2023-STG grant \#101113826) and by the Italian MUR (PRIN 2022 grant \#2022YKTMK3).

\quad

\textcolor{red}{Article published in the \textit{Proceedings of the National Academy of Sciences 121(20) -- PNAS}. DOI: \href{http://doi.org/10.1073/pnas.2307038121}{10.1073/pnas.2307038121}. Please, cite the published version.}
 
\bibliographystyle{elsarticle-num}
\bibliography{references}

\end{document}